\documentclass[preprint]{elsarticle}
\pdfoutput=1 
\usepackage[english]{babel}
\usepackage{lineno}
\usepackage{graphicx,xcolor,layouts}
\usepackage[labelformat=simple]{subcaption}
\usepackage{physics,amssymb,amsmath,mathrsfs,tabularx,xfrac}
\usepackage{url,verbatim,booktabs,epstopdf}
\usepackage{tikz}
\usetikzlibrary{shapes,calc}
\modulolinenumbers[5]

\newcommand{\ml}[1]{{\color{black}#1}}

\journal{International Journal of Heat and Fluid Flow}

\biboptions{sort&compress}
\newcommand{\bfu}{\boldsymbol{u}} 
\newcommand{\bbfu}{\langle\boldsymbol{u}\rangle}  
\newcommand{\bfK}{\mathbf{K}}       
\newcommand{\bfk}{{\boldsymbol{\kappa}}}       
\newcommand{\Ret}{Re_{\tau}}        
\newcommand{\overbar}[1]{\mkern 1.5mu\overline{\mkern-1.5mu#1\mkern-1.5mu}\mkern 1.5mu}
\newcommand{\kxps}{k_{xx}^+}
\newcommand{\kzps}{\smash{k_{zz}^+}\vphantom{k_{xx}^+}}
\newcommand{\kyps}{\smash{k_{yy}^+}\vphantom{k_{xx}^+}}
\newcommand{\utaup}{u_\tau^p}
\newcommand{\utaus}{u_\tau^{s\vphantom{p}}}
\newcommand{\utaut}{u_\tau^{t\vphantom{p}}}

\newcommand{\bcircle}{\raisebox{-2.5pt}{\tikz{
			\draw[line width=1pt,color=white] (-1pt,-4pt) rectangle (1pt,4pt);
			\draw[line width=1pt,fill=black] (0,0) circle (1pt);}}\vphantom{A}}	
\newcommand{\bdash}{\raisebox{-2.5pt}{\tikz{
			\draw[line width=1pt,color=white] (-1pt,-4pt) rectangle (1pt,4pt);
			\draw[dashed,line width=1pt,fill=black] (-0.15,0)--(0.15,0);}}\vphantom{A}}	
\newcommand{\bline}{\raisebox{-2.5pt}{\tikz{
			\draw[line width=1pt,color=white] (-1pt,-4pt) rectangle (1pt,4pt);
			\draw[line width=1pt,fill=black] (-0.15,0)--(0.15,0);}}\vphantom{A}}

\begin{document}
	
\begin{frontmatter}
		
	\title{Resolvent-based design and experimental testing of porous materials for passive turbulence control}
	\author{Andrew Chavarin\fnref{myfootnote}}
	\author{Christoph Efstathiou\fnref{myfootnote}}
	\author{Shilpa Vijay\fnref{myfootnote}}
	\author{Mitul Luhar\fnref{myfootnote}}
	\address{Department of Aerospace and Mechanical Engineering, University of Southern California, Los Angeles, CA}
		
	\begin{abstract}
	An extended version of the resolvent formulation is used to evaluate the use of anisotropic porous materials as passive flow control devices for turbulent channel flow. The effect of these porous substrates is introduced into the governing equations via a generalized version of Darcy's law.  Model predictions show that materials with high streamwise permeability and low wall-normal permeability ($\phi_{xy} = k_{xx}/k_{yy}\gg1$) can suppress resolvent modes resembling the energetic near-wall cycle.  Based on these predictions, two anisotropic porous substrates with $\phi_{xy} > 1$ and $\phi_{xy} < 1$ were designed and fabricated for experiments in a benchtop water channel experiment. Particle Image Velocimetry (PIV) measurements were used to compute mean turbulence statistics and to educe coherent structure via snapshot Proper Orthogonal Decomposition (POD). Friction velocity estimates based on the Reynolds shear stress profiles do not show evidence of discernible friction reduction (or increase) over the streamwise-preferential substrate with $\phi_{xy}>1$ relative to a smooth wall flow at identical bulk Reynolds number. A significant increase in friction is observed over the substrate with $\phi_{xy} < 1$.  This increase in friction is linked to the emergence of spanwise rollers resembling Kelvin-Helmholtz vortices. Coherent structures extracted via POD analysis show qualitative agreement with model predictions.
	\end{abstract}
		
		
\end{frontmatter}
	
\section{Introduction}

\subsection{Motivation}


Functional surfaces such as sharkskin-inspired riblets are some of the simplest and most effective control techniques tested thus far for turbulent friction reduction. Appropriately shaped and sized riblets have shown the ability to reduce drag up to 10\% in laboratory experiments and up to 2\% in real world conditions \citep{luchini1991resistance, robert1992drag,walsh1984optimization,garcia2011hydrodynamic}. 
It is generally accepted that the drag-reducing ability of such surfaces arises from their anisotropy: they offer much less resistance to streamwise flows compared to spanwise flows \citep{luchini1991resistance}. The mean flow in the streamwise ($x$) direction is essentially unimpeded within the riblet grooves, generating high interfacial slip. However, cross-flows in the wall-normal ($y$) and spanwise ($z$) directions arising from turbulence are blocked by the riblets and pushed further from the wall. This blocking effect weakens the quasi-streamwise vortices associated with the energetic near-wall (NW) cycle \citep{robinson1991coherent,smits2011high}, and reduces turbulent mixing and momentum transfer above the riblets \citep{choi1993direct}. Skin friction reduction initially increases with increasing riblet spacing and height. However, above a certain size threshold, performance deteriorates dramatically. Early studies attributed this deterioration of performance to the NW vortices lodging within the riblet grooves \citep{choi1993direct,lee2001flow}.  More recently, \citet{garcia2011hydrodynamic} have shown that a Kelvin-Helmholtz (KH) instability may also contribute to the deterioration of performance.

Recent theoretical efforts and numerical simulations suggest that streamwise-preferential permeable substrates have the potential to reduce drag in wall-bounded turbulent flows through a similar mechanism as riblets \citep{nabil_garcia_dragreduction, rosti2018turbulent,gomez-de-segura_garcia-mayoral_2019}. Simulation results predict that as much as $25\%$ drag reduction may be achievable through the use of anisotropic permeable substrates that have streamwise permeability ($k_{xx}$) that is higher than the wall-normal ($k_{yy}$) or spanwise (\ml{$k_{zz}$}) permeabilities \citep{rosti2018turbulent,gomez-de-segura_garcia-mayoral_2019}. Similar to flow over riblets, a KH instability is also predicted to arise for materials with high wall-normal permeability \citep{nabil_garcia_dragreduction,gomez-de-segura_garcia-mayoral_2019}. However, these predictions remain to be tested in physical experiments.  In this work, we seek to design and fabricate anisotropic porous materials that have the potential to reduce skin friction, and to test these materials in benchtop channel flow experiments.  The design and evaluation of the porous materials is guided by reduced-order models grounded in resolvent analysis \citep{mckeon_sharma_2010_sharma_2010,luhar2014opposition,luhar2015framework,chavarin2020resolvent}, while emerging 3D-printing techniques are used for material fabrication.

\subsection{Previous Theoretical Efforts and Simulations}
As noted above, streamwise-preferential porous materials have the potential to reduce drag in turbulent flows through a similar mechanism as riblets.  With anisotropic porous substrates, high porosity and streamwise permeability contribute to a substantial interfacial slip velocity for the mean flow, while low wall-normal and spanwise permeability limit turbulence penetration into the porous substrate. These effects can also be interpreted in terms of the slip length and virtual origin framework used by \citet{luchini1991resistance} to characterize flows over riblets. Specifically, the interfacial slip velocity for the mean flow can be related to a streamwise slip length $l_U^+$ that determines the virtual origin perceived by the mean flow below the porous interface.  Following standard notation, a superscript $+$ denotes normalization with respect to the friction velocity ($u_\tau$) and kinematic viscosity ($\nu$).  Similarly, the distance to which the turbulent fluctuations penetrate into the porous substrate can be related to a transverse slip length $l_t^+$ that determines the virtual origin for the turbulent cross-flows.  Note that the virtual origin for the turbulent cross-flow can also be interpreted as the location at which the quasi-streamwise NW vortices perceive a non-slipping wall \citep{gomez-de-segura_garcia-mayoral_2019,gm_gs_fairhall_2019}.  The initial decrease in drag over riblets of increasing size has been shown to depend on the difference between the streamwise and transverse slip lengths, $\Delta D \propto l_U^+ - l_t^+$.  Physically, when there is a positive offset between the virtual origins for the mean flow and the transverse fluctuations ($l_U^+ > l_t^+$), the quasi-streamwise NW vortices are pushed into a region of lower mean shear.  This weakens the flow induced by the NW vortices and leads to a reduction in turbulent Reynolds stresses and skin friction.  The virtual origin concept has recently been extended to the case of anisotropic porous substrates.  Specifically, using the Brinkman equations, \citet{nabil_garcia_dragreduction} established the following relationships between the streamwise and transverse slip lengths and permeabilities: $l_U^+ \propto \sqrt{\kxps}$ and $l_t^+ \propto \sqrt{\kzps}$.  These relationships indicate that the initial decrease in drag over anisotropic porous materials is expected to be proportional to the difference between the streamwise and spanwise permeability length scales, $\Delta D \propto l_U^+ - l_t^+ \propto \sqrt{\kxps}-\sqrt{\kzps}$. 

Recent Direct Numerical Simulation (DNS) results obtained by \citet{gomez-de-segura_garcia-mayoral_2019} for turbulent flow over anisotropic permeable substrates show good agreement with the predictions made by \citet{nabil_garcia_dragreduction}.  Specifically, DNS results show the presence of a linear regime in which drag reduction is initially proportional to $\sqrt{\kxps}-\sqrt{\kzps}$.  Thus, drag reduction is expected for substrates with high streamwise permeability and low spanwise permeability.  Further, DNS snapshots of the flow field over drag-reducing porous substrates indicate that the NW dynamics are similar to those observed over smooth walls (i.e., characterized by the presence of elongated streaky structures).  However, there is a decrease in turbulent fluctuation intensity and Reynolds shear stress close to the wall \citep{busse_Sandham_2012, gomez2018turbulent, rosti2018turbulent}.  

These observations indicate that slip length-based models generate useful predictions for the linear drag reduction regime over anisotropic porous substrates.  However, these models do not incorporate the effects of wall-normal permeability, which could also affect turbulence penetration into the substrate.  Moreover, the maximum achievable drag reduction is thought to be limited by the onset of a KH-type instability.  Linear instability analyses predict that the emergence of such instabilities is controlled by $\kyps$ \citep{nabil_garcia_dragreduction,gomez2018turbulent}.  DNS results obtained by \citet{gomez-de-segura_garcia-mayoral_2019} confirm the emergence of energetic spanwise rollers as the wall-normal permeability increases beyond $\sqrt{\kyps} \approx 0.4$.  Momentum balance arguments show that the additional Reynolds shear stress generated by these rollers is responsible for the deterioration of drag reduction performance and, eventually, an increase in drag over the porous substrates. For completeness, we note that spanwise rollers have also been documented in other simulations over permeable walls and isotropic porous substrates \citep{jimenez_pinelli_1999,breugem2006influence,busse_Sandham_2012,rosti2015direct,kuwata2016lattice,kuwata2017direct,rosti2018turbulent}.  Finally, recent geometry-resolving simulations carried out using the lattice Boltzmann method have evaluated the effect of porous materials comprising stacks of perforated plates with permeability ratios as high as $\phi_{xy} = 24$ on a turbulent channel flow. However, these simulations consider materials with $k_{xx}=k_{zz}$ which are not expected to yield drag reduction per the slip length models discussed above.  Indeed, these simulations do not show drag reduction and again confirm that the wall-normal permeability plays a key role in dictating the emergence of spanwise coherent rollers that substantially modify turbulence statistics \citep{kuwata2019extensive}.

\ml{In summary, }the DNS results of \citet{gomez-de-segura_garcia-mayoral_2019} suggest that drag reduction over anisotropic permeable substrates depends on two key factors: the ability of the substrate to weaken the energetic NW cycle, which depends on $\kxps$ and $\kzps$, and the emergence of KH rollers, which is dictated by $\kyps$.  However, these predictions remain to be tested in physical experiments.

\subsection{Previous Experiments}
Previous experimental studies of turbulent flow over porous materials have focused primarily on granular media such as packed beds of spheres \citep[e.g.,][]{zagni1976channel,pokrajac2009velocity,horton2009onset,kim2016experimental} or commercially-available materials such as reticulated foams \citep{suga2010effects,manes2011turbulent,efstathiou2018mean}. These studies have provided significant insight into how porous substrates modify the near-wall turbulent mean flow and statistics.  For instance, they have characterized the effect of substrate permeability on the interfacial slip velocity and the logarithmic region of the mean flow reasonably well, and documented the emergence of spanwise rollers.  Recent work by \citet{kim2020experimental} also attempts to systematically delineate the effect of porosity and interfacial roughness.  
 
However, sphere beds and reticulated foams are approximately isotropic. Few studies have explicitly considered the effect of anisotropic porous materials on turbulent flows.  Even fewer studies have considered the effect of streamwise-preferential materials that have the potential to reduce drag. The early experiments of \citet{kong1982turbulent} over mesh and perforated sheets considered materials with significantly higher wall-normal permeability compared to streamwise permeability, i.e., materials with anisotropy ratio $\phi_{xy} = k_{xx}/k_{yy} < 1$.  Similarly, recent experiments by \citet{suga2018anisotropic} have also focused on materials with $\phi_{xy}<1$. These experiments considered channel flow at bulk Reynolds numbers $Re_b = 900 - 13 600$, where one wall was lined with a porous material.  The porous substrates consisted of layers of co-polymer nets and the resulting anisotropy ratio for these substrates was $\phi_{xy} =  1/190 - 1/1.5$.  In all the cases considered by \citet{suga2018anisotropic}, an increase in friction velocity was observed at the porous wall, and the total friction drag increased by 13-73$\%$. More recent PIV measurements made by \citet{suga2020characteristics} have considered the effect of porous materials with anisotropy ratios $\phi_{xy} \approx 1.25$ and $\phi_{xy} \approx 0.13$ and $k_{xx}=k_{zz}$ on turbulent flow in a square duct. These experiments show broadly similar changes to turbulence statistics and flow structure when compared to measurements made in channel flows with larger aspect ratios.

To the best of our knowledge, the only prior experimental evidence of turbulent drag reduction over porous materials comes from the seal fur tests pursued by \citet{itoh2006turbulent}.  Given the streamwise-preferential nature of sea fur, these observations provide limited support for the theoretical predictions and simulation results discussed in the previous subsection.  However, further verification of these prior results requires a more complete characterization of how streamwise-preferential materials with \textit{known} permeability affect turbulent flows.  

Finally, note that canopies of terrestrial or aquatic vegetation and corals reefs can also be considered anisotropic porous materials.  However, since such substrates do not exhibit high streamwise permeability, the extensive literature on flows over vegetation canopies and coral reefs is not reviewed here for brevity.

\subsection{Contribution and Outline}
In this paper we seek to design and fabricate anisotropic porous materials that have the potential to passively control turbulent flows, and to test these materials in laboratory experiments.  In particular, we leverage advances in additive manufacturing (3D-printing) to fabricate cellular porous materials that have desirable anisotropy ratios $\phi_{xy}>1$, and test the effect of these materials in benchtop channel flow experiments. Measurements made over the material with $\phi_{xy}>1$ are compared against measurements made over a geometrically-similar porous material with $\phi_{xy}<1$ as well as a solid smooth wall.

In order to evaluate the viability of these substrates for passive drag reduction, we extend the resolvent analysis framework the resolvent framework of \citet{mckeon_sharma_2010_sharma_2010} to account for porous substrates. Recent work by \citet{chavarin2020resolvent} shows that the resolvent framework can serve as a useful assessment and design tool for riblets.  In particular, \citet{chavarin2020resolvent} show that the resolvent framework can: (i) predict whether riblets of specified geometry are likely to suppress or amplify the energetic NW cycle, and (ii) be used to test for the emergence of spanwise rollers resembling KH vortices. Here, we use resolvent analysis as a preliminary design and analysis tool: we use it to evaluate the effect of anisotropic porous materials with known permeability on the NW cycle, and to test for the emergence of KH rollers. 

Note that the 3D-printed material with $\phi_{xy}>1$ tested in the experiments did not have sufficiently low wall-normal permeability ($\sqrt{\kyps} \lesssim 0.4$; \citep{gomez-de-segura_garcia-mayoral_2019}) to rule out the emergence of KH rollers. Nevertheless, our measurements show that the material with $\phi_{xy}>1$ leads to minimal changes in friction relative to smooth wall conditions, while the material with $\phi_{xy}<1$ leads to a substantial increase in friction.  These observations confirm that streamwise-preferential materials with $\phi_{xy}>1$ remain promising candidates for passive drag reduction in turbulent flows. Importantly, these experiments also provide preliminary insights into how anisotropic materials with $\phi_{xy}<1$ and $\phi_{xy}>1$ modify the near-wall flow physics.

The remainder of this paper is structured as follows. The resolvent-based modeling framework is described further in \S\ref{sec:modeling}.  The experimental methods are presented in \S\ref{sec:experiments}.  Model predictions and experimental results are discussed together in \S\ref{sec:results}.  Specifically, the resolvent-based predictions used to evaluate the porous materials tested in the experiments are shown in \S\ref{sec:model_predictions}.  Experimental measurements for the mean profile and turbulence statistics are discussed in \S\ref{sec:exp_results} and the flow features identified via snapshot proper orthogonal decomposition (POD) are shown in \S\ref{sec:POD}. One of the inputs required for resolvent analysis is an estimate of the turbulent mean profile.  The model predictions shown in \S\ref{sec:model_predictions} are obtained using a synthetic mean profile computed using an eddy viscosity formulation.  These predictions are compared against those made using the mean profiles \textit{measured} in the experiments in \S\ref{sec:mean_profile}.  Brief concluding remarks are presented in \S\ref{sec:conclusion}. 
	
\section{Modeling}\label{sec:modeling}
In this section, we describe the extension to the resolvent framework to account for porous substrates and provide details on numerical implementation. 

\subsection{Extended Resolvent Formulation}\label{sec:extended_resolvent}
We utilize a modified version of the resolvent formulation proposed by \citet{mckeon_sharma_2010_sharma_2010} to predict the drag performance of anisotropic permeable materials for passive turbulence control. For wall-bounded turbulent flows, the resolvent formulation interprets the Navier-Stokes equations, Fourier transformed in the (approximately) homogeneous streamwise and spanwise directions and in time, as a forcing-response system. For this system the nonlinear convective terms are treated as internal forcing (input) to the system composed from the remaining linear terms of the Navier-Stokes equations. At every wavenumber-frequency combination $\bfk =(\kappa_x,\kappa_z,\omega)$ this internal forcing generates a turbulent velocity and pressure response. A gain-based singular value decomposition of the forcing-response transfer function---the resolvent operator---yields a set of highly amplified velocity and pressure response modes (left singular vectors) and the corresponding forcing-response gains (singular values).  The response modes---termed \textit{resolvent modes}---are flow structures with streamwise and spanwise wavelength $\lambda_x=2\pi/\kappa_x$ and $\lambda_z=2\pi/\kappa_z$, respectively, traveling at speed $c=\omega/\kappa_x$.  The forcing-response gain is a measure of energy amplification in the system, and serves as a metric of control performance. 

Previous work shows that specific high-gain response modes can serve as useful surrogates for energetic structures such as the NW cycle \citep{moarref2013model}. These resolvent modes can therefore serve as building blocks for the design and optimization control strategies \citep{luhar2014opposition,luhar2015framework,nakashima2017assessment,toedtli2019predicting,chavarin2020resolvent}. Specifically, these prior efforts show that suppression of the NW resolvent mode is a useful indicator of drag reduction performance for both active and passive control of wall turbulence.  In other words, if a control technique is unable to suppress the surrogate NW resolvent mode, then it is unlikely to yield drag reduction for the full turbulent flow field.  In addition, recent work by \citet{chavarin2020resolvent} shows that the resolvent framework is also able to predict the emergence of energetic spanwise rollers over riblets that contribute to the deterioration of drag reduction performance \citep{garcia2011hydrodynamic}.  Building on these prior studies, here we use the resolvent framework to test whether a given porous material can (i) suppress the gain for the resolvent mode that serves as a surrogate for the NW cycle, and (ii) limit the emergence of energetic spanwise rollers resembling KH vortices.

For this analysis we formulate the resolvent framework using the volume-averaged Navier-Stokes (VANS) equations in which the effect of anisotropic porous substrates is included via a linear permeability term consistent with Darcy’s law\citep{breugem2006influence}:
\begin{subequations}\label{eqn:governing_VANS} 
	\begin{equation}\begin{gathered}
	\pdv{\bbfu}{t} + \frac{1}{\varepsilon}\grad\cdot\big(\varepsilon\bbfu\bbfu+\varepsilon\boldsymbol{\tau} \big) = \\ -\frac{1}{\varepsilon}\grad(\varepsilon \langle p\rangle) + \frac{1}{\varepsilon\Ret}\grad^2 (\varepsilon\bbfu)-\frac{\varepsilon}{\Ret}\bfK^{-1}\bbfu,
	\end{gathered}\end{equation}
	\begin{equation}
	\grad\cdot(\varepsilon\bbfu) =0
	\end{equation}
\end{subequations}
Here, $\langle \bfu \rangle$ and $\langle p \rangle$ represent the dimensionless volume averaged velocity and pressure respectively, $\varepsilon$ represents the porosity, and ${\bfK}$ is the dimensionless permeability tensor. The equations above have been normalized by the channel height $h$ (see Fig.~\ref{fig:piv-setup}) and the friction velocity $u_{\tau}$ and the friction Reynolds number is given by $\Ret=h^+=u_{\tau}h/\nu$.  The sub-filter scale stresses which arise from volume-averaging the Navier Stokes equations are defined as $\boldsymbol{\tau}=\langle\bfu\bfu\rangle-\bbfu\bbfu$. The unobstructed fluid domain is characterized by porosity $\varepsilon=1$ and infinite permeability.  For this region, the Darcy term drops out of the governing equations, and Eq.~\ref{eqn:governing_VANS} reduces to the standard VANS equations.

Note that the expression above omits the nonlinear Forchheimer term. This assumption was made for several reasons. First, omission of the nonlinear Forchheimer term ensures consistency between the porous medium model used in this paper and the formulations used in prior numerical simulations evaluating the drag-reducing capabilities of anisotropic porous materials \citep{rosti2018turbulent,gomez-de-segura_garcia-mayoral_2019}. Second, including the nonlinear Forchheimer term introduces several additional geometry-dependent parameters as well as a dependence on the pore-scale Reynolds number. This makes interpretation of resolvent-based predictions much more challenging. Further, keep in mind that resolvent analysis probes the linear forcing-response characteristics of the governing equations. As a result, only a linearized version of the nonlinear Forchheimer term would be accounted for in the analysis described below.  In a sense, this is equivalent to simply considering a material with a lower \textit{apparent} permeability \citep{zampogna_bottaro_2016}.

For the remainder of the paper, we focus on porous substrates for which the permeability tensor is diagonal and has the form ${\bfK}= \mathrm{diag}(k_{xx},k_{yy},k_{zz})$.  Further, since we are primarily considering structures (i.e., NW cycle and KH rollers) that are much larger than the pore scale, we assume that the sub-filter scale stresses are negligible.  These assumptions and modeling simplifications are again consistent with those made in recent numerical simulations of flow over anisotropic porous substrates \citep{rosti2018turbulent,gomez-de-segura_garcia-mayoral_2019}.  
To further simplify the expressions in Eq.~\ref{eqn:governing_VANS}, we assume that the porous substrate is spatially homogeneous and has constant porosity. This yields:
\begin{subequations}\label{eqn:governing_VANS2} 
	\begin{equation}\begin{gathered}
	\pdv{\bfu}{t} + \grad\cdot\big(\bfu\bfu) =\\ -\grad p +\frac{1}{\Ret}\grad^2 \bfu-\frac{1}{\Ret}\varepsilon\bfK^{-1}\bfu,
	\end{gathered}\end{equation}
	\begin{equation}
	\grad\cdot \bfu=0,
	\end{equation}
\end{subequations} 
where the $\langle\cdot\rangle$ notation has been omitted for simplicity. 
Resolvent analysis proceeds as follows. First, we employ a standard Reynolds-averaging procedure such that velocity is decomposed into a mean component ($\boldsymbol{U}$) and a fluctuation about this mean ($\bfu^\prime$).  Next, the governing equations for the fluctuations are Fourier-transformed and expressed as
\begin{equation}
    \left[ \begin{array}{c} \bfu_\bfk \\ p_\bfk \end{array}\right] = H_\bfk {\bf f}_\bfk.
\end{equation}
Here, $\bfu_\bfk$ and $p_\bfk$ represents the Fourier-transformed velocity and pressure fluctuations, ${\bf f}_\bfk$ represents the nonlinear forcing terms, and $\boldsymbol{\mathcal{H}}_\bfk$ is the resolvent operator representing the linear forcing-response dynamics. At every wavenumber-frequency combination $\bfk$, an SVD of the discretized resolvent operator, i.e.,
\begin{equation}
    \boldsymbol{\mathcal{H}}_{\bfk}= \sum_{m} \psi_{\bfk,m} \sigma_{\bfk,m} \phi^{\ast}_{\bfk,m},
\end{equation} 
yields forcing modes (right-singular vectors, $\phi_{\bfk,m}$) and velocity/pressure response modes (left-singular vectors, $\psi_{\bfk,m}$) that are ordered based on their forcing-response gain (singular values, $\sigma_{\bfk,m}$). For our analysis, the resolvent operator is scaled to enforce an $L_2$ energy norm and so the change in singular value relative to the smooth wall case can be interpreted as a measure of energy amplification or suppression.  Importantly, previous work shows that the resolvent operator tends to be low-rank at wavenumber-frequency combinations that are energetic in turbulent flows \citep{moarref2013model,mckeon_sharma_2010_sharma_2010}.  Consequently the resolvent operator can be well approximated using a rank-1 truncation after the SVD, i.e., by only considering the first singular values, $\sigma_{\bfk,1}$, and response modes, $\psi_{\bfk,1}$.  \citet{chavarin2020resolvent} show that resolvent analysis with this rank-1 approximation provides useful insight into the effect of riblets on wall turbulence. We retain the rank-1 approximation here as well, and drop the additional subscript 1. For further discussion pertaining to resolvent analysis for wall-bounded turbulent flows and the rank-1 approximation, the reader is referred to several recent studies in this area \citep{mckeon_sharma_2010_sharma_2010,luhar2014structure,mckeon2017engine}.

	
For the remainder of this work, we focus on modes that serve as a surrogate model for the dynamically important NW cycle, i.e., \ml{modes with length and velocity scales that are representative of the streaks and streamwise vortices associated with the NW cycle, $\lambda_x^+ \approx 10^3$, $\lambda_z^+ \approx 10^2$, and $c^+ \approx 10$ \citep[e.g.,][]{robinson1991coherent,mckeon_sharma_2010_sharma_2010,luhar2014opposition}}.  We consider the highest singular value as a measure of performance.  If the singular value---or gain---is reduced over the porous material ($\sigma_{\bfk,p}$) relative to the smooth wall value ($\sigma_{\bfk,s}$), the porous material is likely to suppress the corresponding flow structure.  We also test for the emergence of high-gain spanwise-constant rollers resembling Kelvin-Helmholtz vortices (i.e., with $\kappa_z = 0$) at the porous interface.
	
\subsection{Numerical Implementation}\label{sec:implementation}

\renewcommand\thesubfigure{(\alph{subfigure})}
\begin{figure*}[ht]

	\centering
	\begin{subfigure}[t]{1.0\textwidth}
        \centering
        \includegraphics[width=1.0\textwidth]{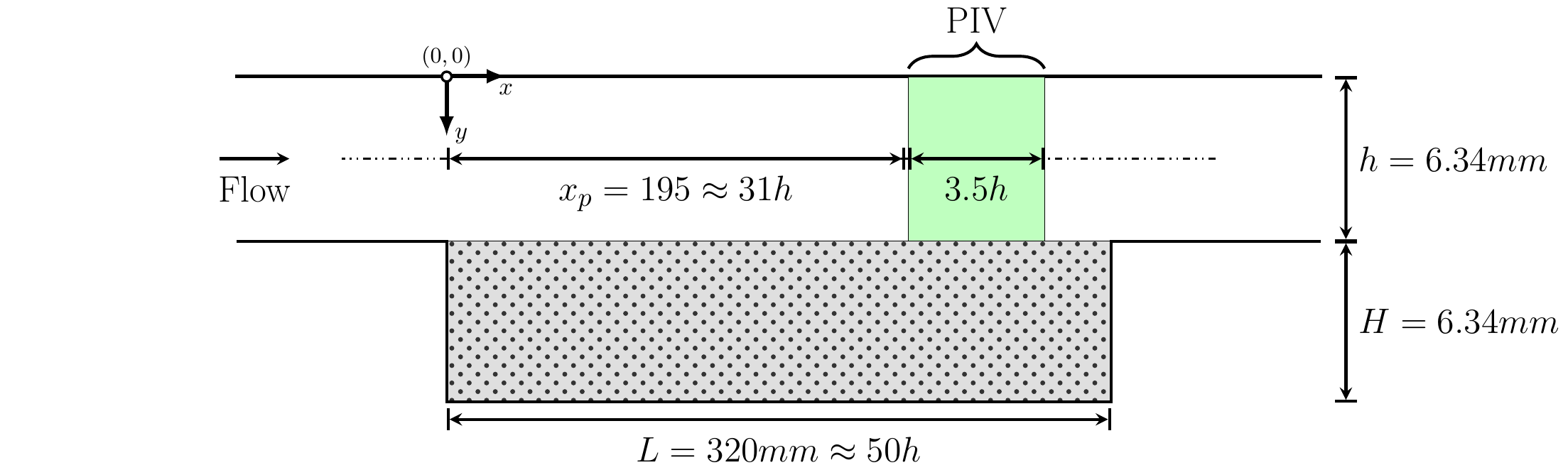} 
        \caption{channel configuration}
	\end{subfigure}
	\vspace{0.5cm}
	
	\begin{subfigure}[t]{0.3\textwidth}
	    \centering
        \includegraphics[height=0.65\textwidth]{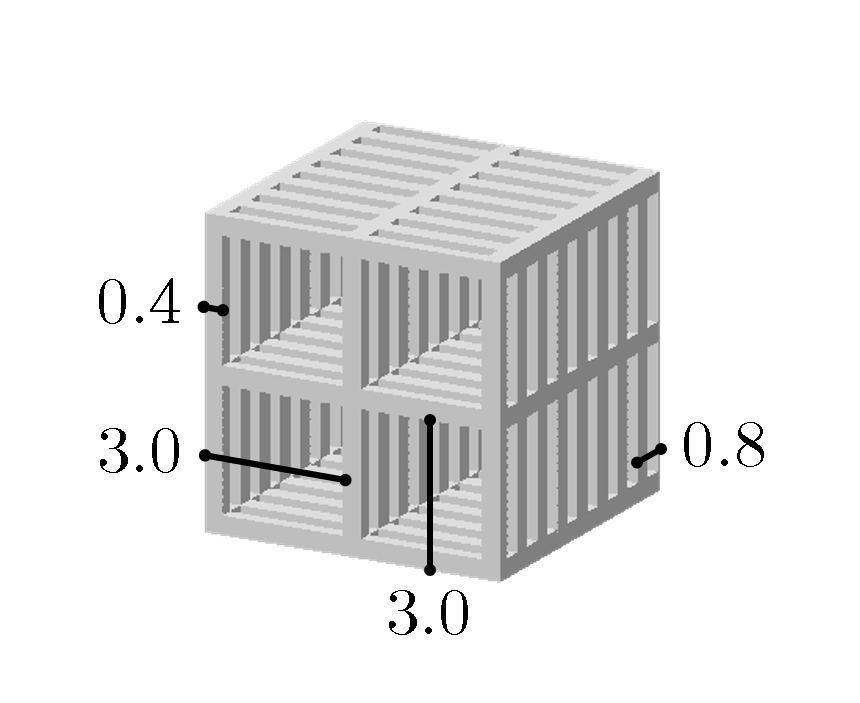} 
        \caption{x-permeable lattice}
	\end{subfigure}
	\begin{subfigure}[t]{0.3\textwidth}
	    \centering
        \includegraphics[height=0.65\textwidth]{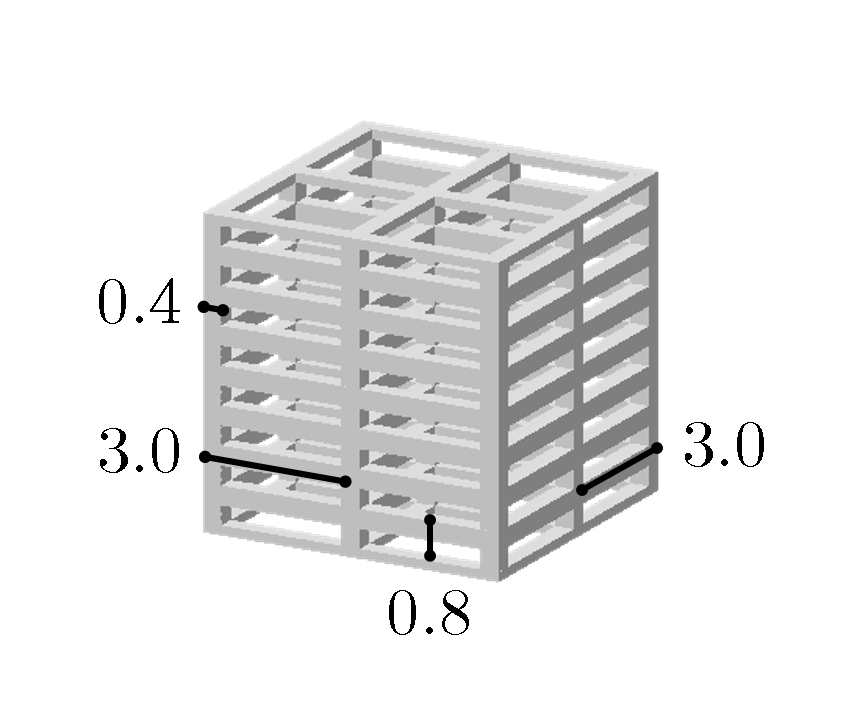} 
        \caption{y-permeable lattice}
	\end{subfigure}
	\begin{subfigure}[t]{0.3\textwidth}
	    \centering
        \includegraphics[height=0.60\textwidth]{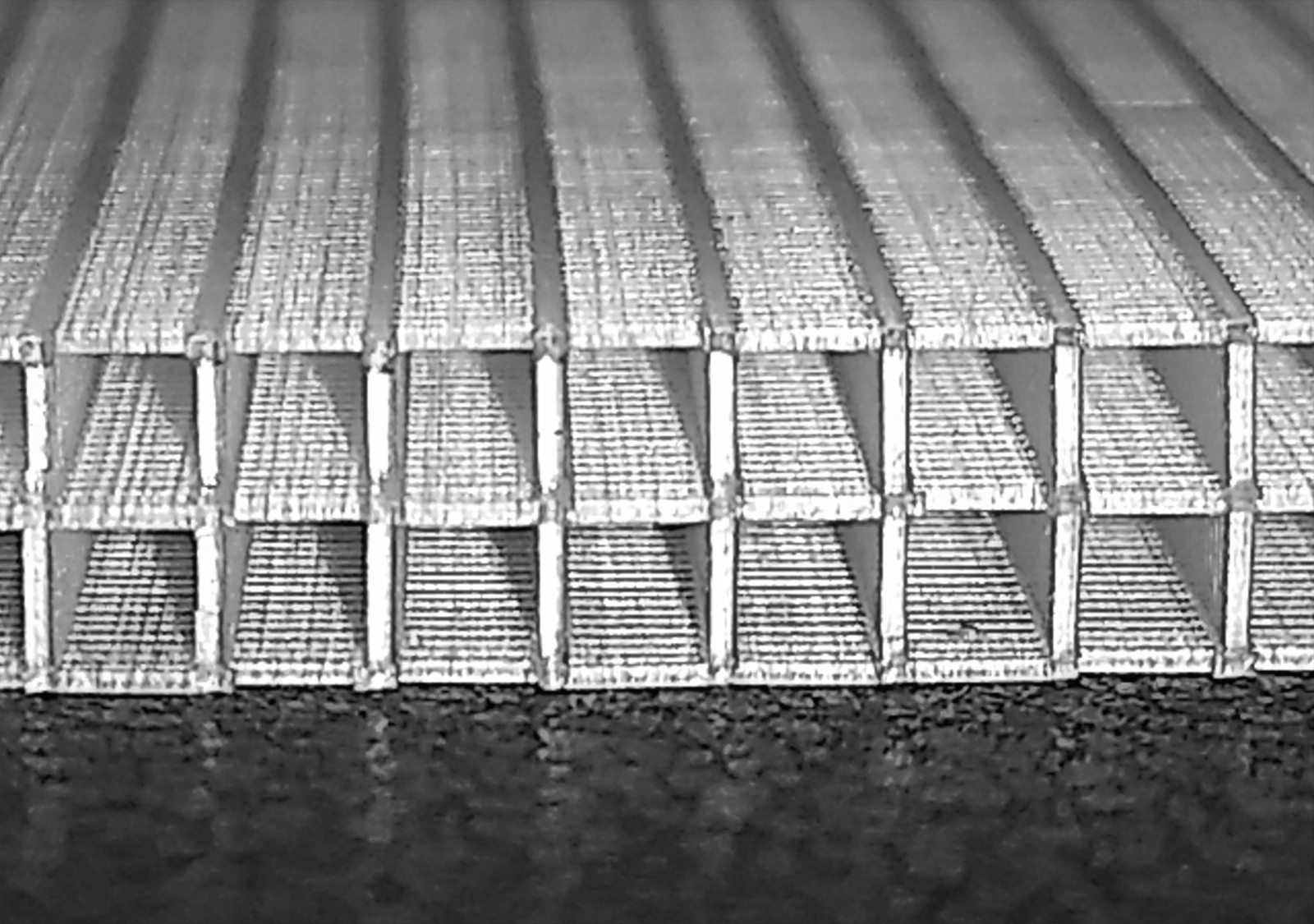} 
        \caption{x-permeable material}
	\end{subfigure}
	
	\caption{(a) Schematic of the channel flow experiment (not to scale). (b,c) Renderings for the x-permeable and y-permeable material showing a cube of dimension 6.34 mm. \ml{All inset dimensions are in mm}. (d) Image of 3D-printed x-permeable porous tile. The height of the tile is 6.34 mm\ml{; for rod sizes and spacings, see panel (b).} Note that the x-permeable material features large openings normal to the streamwise incoming flow. The y-permeable material has large openings in the wall-normal direction.} 
	\label{fig:piv-setup}
\end{figure*}

Previous theoretical and numerical efforts have primarily considered a symmetric channel geometry, with porous materials at both the upper and lower walls \citep{rosti2018turbulent,gomez-de-segura_garcia-mayoral_2019}.  However, this geometry would have limited optical access for Particle Image Velocimetry (PIV) in the laboratory experiments discussed below. Instead, we consider an asymmetric channel geometry corresponding to the experimental setup shown in Fig.~\ref{fig:piv-setup}(a). The unobstructed region of the channel spans $y \in [0,h]$ and the porous material occupies the region corresponding to $y\in(h,H+h)$, with $H = h$.  We generate model predictions for $\Ret = u_\tau h/\nu = 120$, which corresponds roughly to the smooth wall conditions tested in the experiments, and for $\Ret = 360$, which corresponds to prior numerical simulations \citep{rosti2018turbulent}. As noted earlier, the spatially-homogeneous porous substrate is defined by its principal permeability components ${\bfK} = \mathrm{diag}(k_{xx},k_{yy},k_{zz})$.  For simplicity, the porosity is set to $\varepsilon = 1$, though the materials tested in the experiment have a porosity \ml{of} $\varepsilon \approx 0.83$. 

No-slip boundary conditions are applied at the true walls, which are located at $y=0$ and $y=H+h$. The interface between the porous substrate and the unobstructed domain is located at $y=h$. The resolvent operator is discretized in the wall-normal direction using the Chebyshev collocation method detailed by \citet{rectangular_block_operators_trefethen}. This approach allows us to discretize the unobstructed and porous domains independently and couple these two domains through simplified versions of the jump boundary conditions proposed by \citet{ochoa1995momentum1}. The following boundary conditions are applied at the interface:
\begin{subequations}
	\begin{equation}
	\left.\bfu\right\lvert_{y=h^-} =\left.\bfu\right\rvert_{y=h^+},
	\end{equation}
	\begin{equation}
	\left.{p}\right\lvert_{y=h^-} =\left.{p}\right\lvert_{y=h^+},
	\end{equation}
	\begin{equation}
	\left.\pdv{u}{y}\right\lvert_{y=h^-} - \frac{1}{\varepsilon} \left.\pdv{u}{y}\right\lvert_{y=h^+}=0,\\
	\end{equation}
	\begin{equation}
	\left.\pdv{w}{y}\right\lvert_{y=h^-} - \frac{1}{\varepsilon} \left.\pdv{w}{y}\right\lvert_{y=h^+}=0,
	\end{equation}
\end{subequations}
where $h^-$ and $h^+$ refer to $y$-locations on either side of the interface. Similar formulations have been used in prior stability analyses \citep{tilton_cortelezzi_2006,tilton_cortelezzi_2008} as well as numerical simulations of turbulent flows over porous materials \citep{rosti2015direct}.

For the model predictions presented in this paper, a total of 226 Chebyshev nodes are used for discretization in the wall-normal direction. The nodes are divided evenly between the porous and the unobstructed domain. Further grid refinement beyond this point led to changes in singular values smaller than $O(10^{-4})$.  Note that several prior efforts studying the effect of porous media on turbulent flows have employed a continuous formulation in which porous medium properties (e.g., $\varepsilon$, $\bfK$) vary smoothly from values prescribed in the homogeneous region to the unobstructed fluid flow over a finite transition zone \citep{breugem2006influence}.  This continuous approach was considered for the present study as well, but it led to poorer convergence in singular values for resolvent analysis. Moreover, convergence properties were strongly dependent on the size of the transition zone.  Even with a relatively large transition region (i.e., greater smoothing), convergence in singular values to $O(10^{-4})$ typically required more than $N = 500$ nodes in the wall-normal direction.  Since this continuous approach involved an additional tuning parameter (i.e., the size of the transition zone) and led to poorer convergence, the split-domain approach was preferred.

Construction of the resolvent operator also requires a mean velocity profile, $\boldsymbol{U}=[\overbar{U}(y),0,0]$. The mean velocity is predicted from the Reynolds-averaged mean flow equation, with the Reynolds stress term modeled using an eddy viscosity. 
For the unobstructed region, the eddy viscosity profiles for our analysis are generated using the analytical model developed by \citet{reynolds_tiederman_1967} and the linear permeability term is set to zero.  For the porous layer, the eddy viscosity is set to zero and the permeability is set to $\bfK$. We recognize that the eddy viscosity formulation of \citet{reynolds_tiederman_1967} was developed for smooth walls and may not be appropriate over permeable substrates. Ideally, we would use an improved eddy viscosity model that accounts for turbulence in the interfacial region as well as the effects of spanwise and wall-normal permeability. However, since there is currently insufficient information in the literature to develop a reliable eddy viscosity model over anisotropic porous materials, we use the standard smooth wall profile to create a purely \textit{predictive} modeling framework. We also note that a similar approach led to useful resolvent-based predictive models for riblets \citep{chavarin2020resolvent}.

Since resolvent-based predictions are sensitive to the exact form of the mean profile, in \S\ref{sec:mean_profile}, we compare model predictions obtained using the \textit{synthetic} mean profile discussed above against predictions obtained using a profile fitted to the experimental measurements described below. This \textit{fitted} profile is synthesized from the experimental measurements as follows.  First, the mean velocity ($\overbar{U}(y)$) and Reynolds shear stress ($=-\overbar{u^\prime v^\prime}$) profiles are obtained from the PIV measurements described below by averaging in time and in the streamwise direction.  These profiles are then used to estimate the eddy viscosity profile, $\nu_T=-\overbar{u^\prime v^\prime}/\left(d\overbar{U}/dy\right)$, where the mean shear is $d\overbar{U}/dy$ is approximated using a finite difference scheme. 
The points near the maximum in the mean profile, corresponding to $d\overbar{U}/dy \approx 0$, are removed and a smooth cubic spline is fitted to the resulting profile.  Finally, the eddy viscosity profile is allowed to smoothly transition to zero in the porous medium and values for which $\nu_t(y)<0$ are removed and set to $0$.  This fitted eddy viscosity profile is then used to generate predictions for the mean profile (see Fig.~\ref{fig:mean_vel_comparison}) used in the resolvent operator. 


	
	%

\section{Experimental Methods}\label{sec:experiments}

\subsection{3D Printed Porous Materials}\label{sec:3dprint}

For this study, two custom anisotropic porous materials were fabricated using a stereo-lithographic 3D printer (formlabs Form 2) based on input from the resolvent-based predictions described below. The porous material microstructure consisted of a cubic lattice of rectangular rods with constant cross-section and varying spacing in the $x$, $y$, and $z$ directions. Fabrication constraints (printing resolution, allowable unsupported lengths, resin drainage) limited the maximum anisotropy that could be achieved. The anisotropy was varied by controlling the size of the pores normal to the spanwise and streamwise directions. The minimum pore size was dictated by the printer resolution as the rods fused and the surface became solid if the separation between two rods fell below the minimum resolution. 
Moreover, the maximum pore size was limited by the maximum overhang between rods because with excessive overhang, the horizontal rod sagged and deviated from the design geometry. Considering these limitations, the following two geometries represented a good compromise between reliable fabrication and anisotropy. The first case with spacings $s_x = 0.8$ mm and $s_y=s_z=3.0$ mm contained larger pores facing the streamwise direction and small pores facing the wall-normal and spanwise directions (see Fig.~\ref{fig:piv-setup}(b)).  The second geometrically similar, but rotated, case with $s_x = s_z = 3.0$ mm and $s_y = 0.8$ mm contained larger pores facing the wall-normal direction (see Fig.~\ref{fig:piv-setup}(c)). These materials are referred to as \textit{x-permeable} and \textit{y-permeable}, respectively, for the remainder of this paper. For both geometries, the rod cross-section was a square of size $d \times d$, with $d=0.4$ mm.  The porosity of the designed geometry was $\varepsilon \approx 0.87$. A 3D-printed sample of the x-permeable material is shown in Fig.~\ref{fig:piv-setup}(d). Porosity for the 3D-printed material was estimated based on weight measurements of a cubic sample and the known density of the resin used for printing.  These measurements indicated a slightly-lower porosity of $\varepsilon \approx 0.83 \pm 0.01$ for the as-printed material. This suggests that some of the smaller internal pores may have been clogged.

Following the approach of \citet{zampogna_bottaro_2016}, the permeability tensor ($\bfK$) for the designed anisotropic porous materials was estimated by solving independent forced Stokes flow problems for a unit cell of the cubic lattice in the ANSYS Fluent software package.  Due to the symmetric nature of the microstructures tested, only two Stokes flow problems were required to determine the permeability components $k_{xx}$ and $k_{yy}$; $k_{zz}$ is equal to either $k_{xx}$ or $k_{yy}$ depending on configuration (i.e., x-permeable or y-permeable). For these two Stokes flow problems, a uniform body forcing of unit amplitude was applied in the direction of the permeability component being evaluated. The permeability was determined from the resulting volume-averaged velocity using Darcy's law. Periodic boundary conditions were applied to the boundaries of the unit cell and a no-slip condition was applied at the solid boundaries. A time-marching scheme was used for each of these simulations. The solutions were determined to be at steady state when the residual in the permeability was less than $10^{-6}$. A mesh independence study confirmed that our results were grid converged. The resulting permeability estimates are shown in Table~\ref{table:perm_table}. The anisotropy ratio based on these numerical permeability estimates is $\phi_{xy} = k_{xx}/k_{yy} \approx 8$ for the x-permeable case and $\phi_{xy} \approx 1/8$ for the y-permeable case.

The permeability of the 3D-printed material was also estimated based on laboratory pressure drop measurements.  Specifically, a custom-designed square duct was developed to hold 3D-printed cubes of the porous material. A submersible pump placed in a larger reservoir was used to drive flow across these samples. The flow rate was controlled using an electronic proportioning valve (Omega PV14 series) and the pressure drop across the sample was measured using a differential pressure transducer (Omega PX409 series).  Permeabilities in the directions of the small and large pores were estimated by fitting Darcy's law, including a Forchheimer correction term, to the velocity-pressure drop measurements. The measured permeability values are also shown in Table~\ref{table:perm_table}.  The measured permeability for flow in the direction of the large pores (i.e., $k_{xx}$ for the x-permeable material) was consistent with the numerical predictions, at $93 \pm 8 \%$ of the value obtained in the Stokes flow simulations.  However, measured permeability for flow in the direction of the small pores (i.e., $k_{yy}=k_{zz}$ for the x-permeable material) was significantly lower, at $20 \pm 1\%$ of the value obtained in the simulations.  This further suggests the possibility of clogging in the direction of the small pores. The permeability measurements suggest anisotropy ratios of $\phi_{xy} \approx 36$ for the x-permeable material and $\phi_{xy} \approx 1/36$ for the y-permeable material.

\begin{table}
	\centering
	\newcommand{\tstrut}{\rule{0pt}{2.6ex}}
	\begin{tabular}{ccccc}
	    \toprule
		& $\varepsilon$  & $\sfrac{k_{xx}}{H^2}$ & $\sfrac{k_{yy}}{H^2}$ & $\sfrac{k_{zz}}{H^2}$  \\ \midrule\tstrut
		& \multicolumn{4}{c}{predicted values}\\ \cmidrule(r){2-5}
		x-permeable &   0.87  &  $4.3\times 10^{-3}$  &  $5.5\times 10^{-4}$  &  $5.5\times 10^{-4}$ \\ \tstrut
		y-permeable&   0.87	 &  $5.5\times 10^{-4}$  &  $4.3\times 10^{-3}$  &  $5.5\times 10^{-4}$ \\ \tstrut \tstrut%

		& \multicolumn{4}{c}{measured values}\\ \cmidrule(r){2-5}
		x-permeable &   0.83  &  $4.0\times 10^{-3}$  &  $1.1\times 10^{-4}$  &  $1.1\times 10^{-4}$ \\ \tstrut
		y-permeable&   0.83	 &  $1.1\times 10^{-4}$  &  $4.0\times 10^{-3}$  &  $1.1\times 10^{-4}$\\
		\bottomrule
	\end{tabular}\normalsize
	\caption{Dimensionless permeability estimates for the 3D-printed porous materials.  $H=6.34$ mm is the height of the porous substrates tested in the channel flow experiments.}
	\label{table:perm_table}
\end{table}

Finally, note that the permeability values presented in Table~\ref{table:perm_table} are normalized based on the height of the porous substrates tested in the experiments, $H = 6.34$ mm.  Based on friction velocities estimated from the PIV measurements (described below and presented in Table~\ref{table:exp_table}), we anticipate $H^+ \approx 129$ for the x-permeable material. This means that, in inner-normalized terms, permeabilities estimated from the Stokes flow simulations for the x-permeable material correspond to $(\sqrt{\kxps},\sqrt{\kyps}) \approx (8.5,3.0)$ and permeabilities estimated from pressure drop measurements correspond to $(\sqrt{\kxps},\sqrt{\kyps}) \approx (8.2,1.4)$. Recall that the numerical simulations \citet{gomez-de-segura_garcia-mayoral_2019} indicate that drag reduction performance deteriorates for $\sqrt{\kyps} \gtrsim 0.4$ due to the emergence of spanwise rollers. Thus, even though the x-permeable material has the desired streamwise-preferential anisotropy, it is susceptible to the emergence of spanwise rollers. This possibility is evaluated further in \S\ref{sec:results}.

\subsection{Channel Flow Experiment}\label{sec:experimental_methods}
The anisotropic porous substrates described above were tested in a turbulent channel flow experiment, albeit at very low Reynolds number. A schematic of the experimental setup is shown in Fig.~\ref{fig:piv-setup}. A custom test section was machined from acrylic with a cutout of length $L=320$ mm designed to hold the porous substrates. The width of the test section was $W=50$ mm, and the height of the unobstructed region was $h = 6.34$ mm.  Though the aspect ratio of the channel is relatively low, $W/h = 7.89$, measurements made at the centerline are expected to be representative of channel flows at larger aspect ratios \citep{vinuesa2014aspect,suga2020characteristics}. For instance, DNS results obtained by \citet{vinuesa2014aspect} indicate that the mean velocity profile and turbulence statistics obtained at the centerline of ducts with aspect ratios ranging from 1 to 7 are similar to those obtained in turbulent channel flow simulations when normalized by the local friction velocity. Similarly, recent PIV measurements made by \citet{suga2020characteristics} over porous materials in a square duct show broadly similar mean turbulence statistics compared to measurements made in channels with higher aspect ratios. The cutout was located approximately 150 mm from the \ml{inlet}, and allowed for 3D-printed tiles of thickness $H=h=6.34$ mm to be mounted flush with the smooth wall upstream of the cutout. The number of pores accommodated over the height of the tiles was limited to 2 for the x-permeable case and 8 for the y-permeable case, indicating limited separation between the pore-scale and outer-scale flow. For a baseline comparison, experiments were also carried out with a solid smooth walled insert placed in the cutout. 

We recognize that there is insufficient scale separation between the pore size and the height of the porous medium and, as such, the volume-averaged representation shown in Eq.~\ref{eqn:governing_VANS} is not truly valid. The large pore sizes are driven by the minimum pore size and the desire to generate a maximum anisotropy ($\phi_{xy}=k_{xx}/k_{yy}$). Nevertheless, this represents the first set of experiments over custom-designed, anisotropic porous media.

Flow in the channel was generated using a submersible pump placed in a large water tank.  The flow rate was controlled using an electronic proportioning valve.  The volumetric flow rate was $Q=92$ cm$^3$/s for the smooth wall and x-permeable cases, and $Q=82$ cm$^3$/s for the y-permeable case.  Thus, the bulk Reynolds number was $Re_b = Q/(W\nu) = 1840$ for the smooth wall and x-permeable case, and $Re_b = 1640$ for the y-permeable case.  This corresponds to the lower end of the $Re_b$ ranges considered by \citet{suga2018anisotropic} in recent experiments over anisotropic porous materials.  
We also recognize that the bulk Reynolds numbers considered in this study are towards the low end of what is typically classified as being turbulent.  However, moving to higher $Re_b$ would have either required an increase in channel height or an increase in flow rate.  An increase in channel height would have necessitated increases in channel width and length to maintain aspect ratio, and thereby a substantial increase in 3D-printing requirements.  An increase in flow rate would have led to an increase in friction velocity, and thereby increased $\sqrt{\kyps}$ for the x-permeable material, pushing it further into the drag-increasing regime identified by \citet{gomez-de-segura_garcia-mayoral_2019}.

A 5W continuous wave laser with integrated optics was used to generate a laser sheet in the streamwise-wall normal direction at mid-span. The thickness of the light sheet was approximately 1 mm, and so the PIV measurements discussed below implicitly include some spanwise averaging. A high-speed camera (Phantom VEO-410L) was used to capture images near the downstream end of the porous section. Recent turbulent boundary layer experiments over isotropic porous foams show that the flow adjusts to the new substrate over a streamwise distance of $\approx 30H-40H$, where $H$ is the porous layer thickness \citep{efstathiou2018mean}.  To provide an adequate development length therefore, the PIV field of view began 195mm ($\approx 31H$) from the leading edge of the porous section and extended 22mm ($\approx 3.5H$) downstream. For the smooth wall case, the total development length including the 150 mm section upstream of the cutout was greater than $50h$. Images were acquired at 2kHz for 10 seconds for a total of 20,000 images. The total duration of the measurements is approximately 100 turnover times, where the turnover time is estimated as $(H+h)/U_b$ and the bulk-averaged velocity is defined as $U_b = Q/[W(H+h)]$. The images were processed in PIVlab \citep{thielicke2014pivlab} using the Fast-Fourier transform routine with a minimum box size of 16 pixels and 50\% overlap, which yielded 36 (vertical) x 125 (horizontal) data points in the unobstructed section.  Based on friction velocities computed from the PIV measurements, the vertical resolution was $\Delta y^+ = \Delta x^+ = 3.3-5$ in inner units.  

To test for flow development, turbulence statistics were computed for three different streamwise subsections of the PIV measurements, corresponding to $12.5\%-37.5\%$, $37.5\%-62.5\%$, and $62.5\%-87.5\%$ of the PIV measurement window.  Statistics computed in these different sections showed no significant differences.  Specifically, integrated turbulence intensities for the streamwise and wall-normal velocity fluctuations changed by less than $2\%$ across these three windows.  Moreover, these changes were non-monotonic in the streamwise direction, which suggests that development effects did not play a role.
\ml{In physical terms, the streamwise locations of the three PIV subsections used to test for flow development correspond to roughly $54.8h-55.7h$, $55.7h-56.6h$, and $56.6h-57.5h$ from the channel inlet, or $31.2H-32.1H$, $32.1H-33.0H$, and $33.0H-34.9H$ from the start of the cutout.  Recall that the height of the porous medium ($H$) is identical to the height of the unobstructed channel ($h$). The smooth wall development length ($>50h$) is consistent with guidelines suggested in previous literature \citep[see e.g.,][]{byrne1969turbulent,zanoun2009study} though some studies suggest that subtle changes in the flow continue to occur for over 100 channel heights downstream of the entrance \citep{lien2004entrance,vinuesa2014new}.  There is limited prior work characterizing flow development over anisotropic porous materials.  However, the development length over the porous substrates ($>30H$) exceeds guidelines suggested in previous studies evaluating smooth-to-rough wall transitions \citep{antonia1971response}, flow development over isotropic high-porosity foams \citep{efstathiou2018mean}, as well as flow past backward-facing steps (c.f., the cutout housing the porous material; \citep{le1997direct}).}

\section{Results and Discussion}\label{sec:results}
In this section, we first present resolvent-based predictions for the mode that serves as a surrogate for the NW cycle and for spanwise constant KH-rollers, focusing on the x-permeable and y-permeable cases tested in experiment (\S\ref{sec:model_predictions}). Note that these predictions are generated for a friction Reynolds number corresponding to the baseline smooth wall case, using the synthetic mean velocity profile, and the permeability values estimated from the Stokes flow simulations.  In other words, the predictions presented in \S\ref{sec:model_predictions} only make use of information that is likely to be available prior to material fabrication and testing in laboratory experiments.  We then present experimental measurements for the mean flow, turbulence statistics and flow structure (\S\ref{sec:exp_results}-\S\ref{sec:POD}).  Finally, in \S\ref{sec:mean_profile}, we compare the \textit{a priori} model predictions generated using the synthetic mean profiles and numerical permeability estimates against predictions obtained using mean profiles fitted to experimental data and measured permeabilities for the 3D-printed materials.

\subsection{Model Predictions}\label{sec:model_predictions}

\begin{figure*}[!ht]
	\centering
	\includegraphics[width=0.8\textwidth]{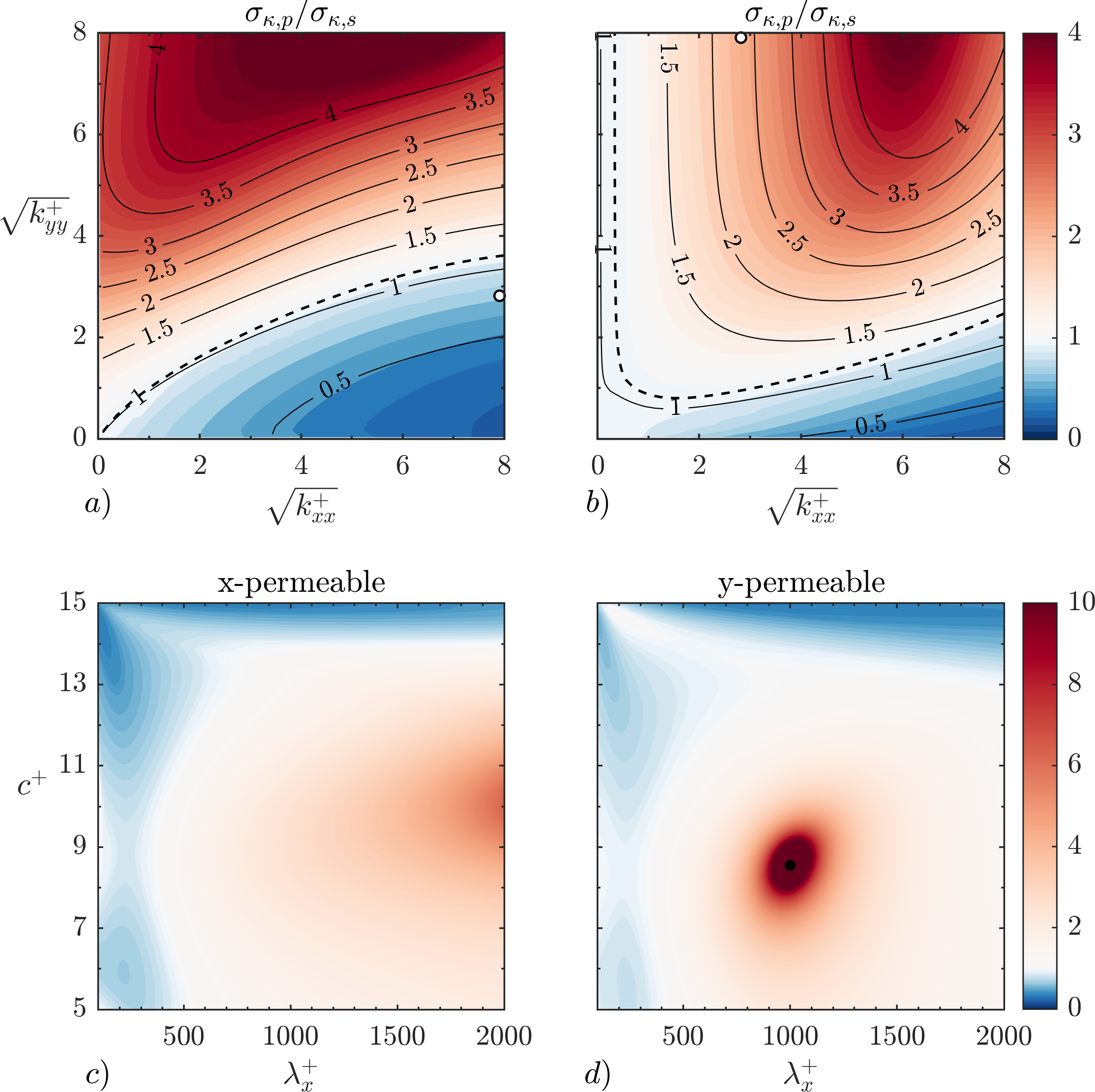}
	\caption{(a,b) Predicted singular value ratios ($\sigma_{\bfk,p}/\sigma_{\bfk,s}$) for resolvent modes resembling the NW cycle as a function of streamwise and wall-normal permeability length-scales.  Color contours show predictions for $\Ret=120$; solid black lines correspond to $\Ret=360$.  Predictions in panel (a) are for substrates which have a similar configuration to x-permeable material ${\bfK} = \mathrm{diag}(k_{xx},k_{yy},k_{zz}=k_{yy})$.  Predictions in panel (b) are for substrates with a similar configuration to the y-permeable substrate, ${\bfK} = \mathrm{diag}(k_{xx},k_{yy},k_{zz}=k_{xx})$.  The ($\protect\circ$) symbols in (a) and (b) correspond roughly to \ml{numerically-predicted} permeabilities for the x-permeable and y-permeable materials tested in the experiments. (c,d) Amplification of spanwise-constant modes at $\Ret=120$ relative to the smooth wall case as a function of streamwise wavelength and mode speed for (c) the x-permeable substrate and (d) for the y-permeable substrate, i.e., for permeability values labeled with ($\protect\circ$) symbols in (a) and (b).} 
	\label{fig:sigma_kh_combined}
\end{figure*}

Figures~\ref{fig:sigma_kh_combined}(a,b) show the predicted change in singular values for resolvent modes resembling the NW cycle (i.e., resolvent modes with $\lambda_x^+ = 10^3$, $\lambda_z^+ = 10^2$, and $c^+ = 10$) over anisotropic porous substrates as a function of their streamwise and wall-normal permeability length scales, $\sqrt{\kxps}$ and $\sqrt{\kyps}$.  Note that the contours show the forcing-response gain for the porous material normalized by the smooth wall value at the same $\Ret$; $\sigma_{\bfk,p}/\sigma_{\bfk,s}<1$ indicates mode suppression and $\sigma_{\bfk,p}/\sigma_{\bfk,s}>1$ indicates mode amplification. For all the predictions shown in Fig.~\ref{fig:sigma_kh_combined} the mean flow was computed using the synthetic eddy viscosity profile. The friction Reynolds number corresponds to the baseline smooth wall case tested in the laboratory experiments, $Re_\tau = u_\tau h/\nu \approx 120$.


Consistent with prior simulation results \citep{rosti2018turbulent,gomez-de-segura_garcia-mayoral_2019}, porous substrates with high streamwise permeability and low wall-normal permeability are found to suppress the NW mode, which is known to be a useful predictor of drag reduction performance \citep{chavarin2020resolvent}. In general, mode suppression increases as the permeability ratio increases, $\phi_{xy} \gg 1$, though there are some subtle differences between the results presented in Fig.~\ref{fig:sigma_kh_combined}(a) for substrates with $\kzps = \kyps$ and in Fig.~\ref{fig:sigma_kh_combined}(b) for substrates with $\kzps = \kxps$.  The substrates shown in panel (a) produce greater mode suppression than those shown in panel (b) for $\phi_{xy} \gg 1$.  This is consistent with the virtual origin model proposed in previous studies \citep{nabil_garcia_dragreduction,gomez-de-segura_garcia-mayoral_2019}, which suggests that turbulence penetration into the porous medium is dictated by the spanwise permeability, and that the initial decrease in drag depends on the difference between the streamwise and spanwise permeability length scales $\Delta D \propto \sqrt{\kxps} - \sqrt{\kzps}$. 

The predictions shown in Figs.~\ref{fig:sigma_kh_combined}(a,b) do not change substantially from $\Ret = 120$ (colored shading) to $\Ret =360$ (solid black lines). In particular, the location of the neutral curve corresponding to $\sigma_p/\sigma_s = 1$ (i.e., no change in gain) is very similar for both Reynolds numbers. 
In addition, the trends in the suppression and amplification of the NW-mode remain the same between $\Ret =120$ and  $\Ret =360$.

For the specific porous materials designed for our experiments, numerical permeability estimates suggest $(\sqrt{\kxps},\sqrt{\kyps})$ =(7.9,2.8) for the x-permeable case and $(\sqrt{\kxps},\sqrt{\kyps})$ =(2.8,7.9) for the y-permeable case at $\Ret = 120$. These values were computed from the dimensionless permeability listed in Table~\ref{table:perm_table} assuming $H^+ =\Ret = 120$. These specific permeability ratios are labeled using $\circ$ markers in Figs.~\ref{fig:sigma_kh_combined}(a,b). Model predictions indicate that the x-permeable substrate suppresses resolvent modes resembling the NW cycle by approximately $25-30\%$ (see $\circ$ marker in Fig.~\ref{fig:sigma_kh_combined}(a)).  In contrast, the y-permeable substrate leads to significant mode amplification, with $\sigma_{\bfk,p}/\sigma_{\bfk,s} > 2$ (see $\circ$ marker in Fig.~\ref{fig:sigma_kh_combined}(b)).  This is broadly consistent with previous simulations, which indicate that drag reduction is only expected over streamwise-preferential materials.  

In addition to the suppression or amplification of the NW cycle, the other factor that controls the drag-reduction performance of anisotropic porous substrates is the emergence of KH rollers \citep{gomez-de-segura_garcia-mayoral_2019, nabil_garcia_dragreduction, breugem2006influence, chandesris2013direct}. Linear stability analysis and simulations suggest that the appearance of such rollers is linked to a relaxation of the wall-normal permeability.  Specifically, the recent simulations of \citet{gomez-de-segura_garcia-mayoral_2019} indicate that the spanwise rollers emerge as the wall normal permeability increases beyond $\sqrt{\kyps} \approx 0.4$.  Unfortunately, due to fabrication constraints, both the x-permeable and y-permeable substrates tested here are expected to have wall-normal permeabilities larger than this threshold value. Figures~\ref{fig:sigma_kh_combined}(c,d) show the normalized gain for spanwise-constant ($\kappa_z=0$) resolvent modes over the designed x-permeable and y-permeable materials, respectively.  For the y-permeable case a region of high amplification is visible in Fig.~\ref{fig:sigma_kh_combined}(d) for structures with streamwise wavelength $\lambda_x^ + \approx 700-1200$ and mode speed $c^+\approx 5 - 10$.  The wave speed of these structures indicates that these structures are localized close the fluid-porous interface. The most amplified structure in this region has $\lambda_x^+\approx 1000$ and $c^+\approx 8.5$.  The gain for this structure increases by a factor of approximately 100 relative to the smooth wall case.  In contrast, Fig.~\ref{fig:sigma_kh_combined}(c) shows that there is no localized maximum in relative amplification for spanwise-constant structures for the x-permeable case. Model predictions indicate that spanwise rollers with $\lambda_x^+ \ge 600$ and $c^+ \approx 4-15$ are amplified relative to the smooth wall case and amplification generally increases with increasing wavelength.  The gain for the most amplified structure in Fig.~\ref{fig:sigma_kh_combined}(c) increases by a factor of approximately 6 relative to the smooth wall case.  Thus, even though the x-permeable material is susceptible to the emergence of spanwise-constant structures, the degree of energy amplification relative to the smooth wall case is more limited compared to that for the y-permeable material.

Together, the predictions presented in section indicate that the x-permeable material is likely to suppress the energetic NW cycle but could give rise spanwise rollers resembling KH vortices.  The y-permeable material is likely to further amplify the NW cycle and give rise to spanwise rollers that are amplified significantly relative to the smooth wall case.  These predictions are compared against the measurements made in the benchtop channel flow experiments in \S\ref{sec:exp_results} and \S\ref{sec:POD} below.  Keep in mind that the predictions made in this section only make use of \textit{a priori} information, i.e., numerical permeability estimates for the designed materials and synthetic mean profiles computed using a smooth-wall eddy viscosity model.  In \S\ref{sec:mean_profile}, we generate additional predictions using mean profiles fitted to the measurements described below and the permeabilities estimated from pressure drop measurements across the 3D-printed materials.

	
\subsection{Mean Flow and Turbulence Statistics}\label{sec:exp_results}
	
\begin{figure*}[!ht]
	\centering
	\includegraphics[width=0.85\textwidth]{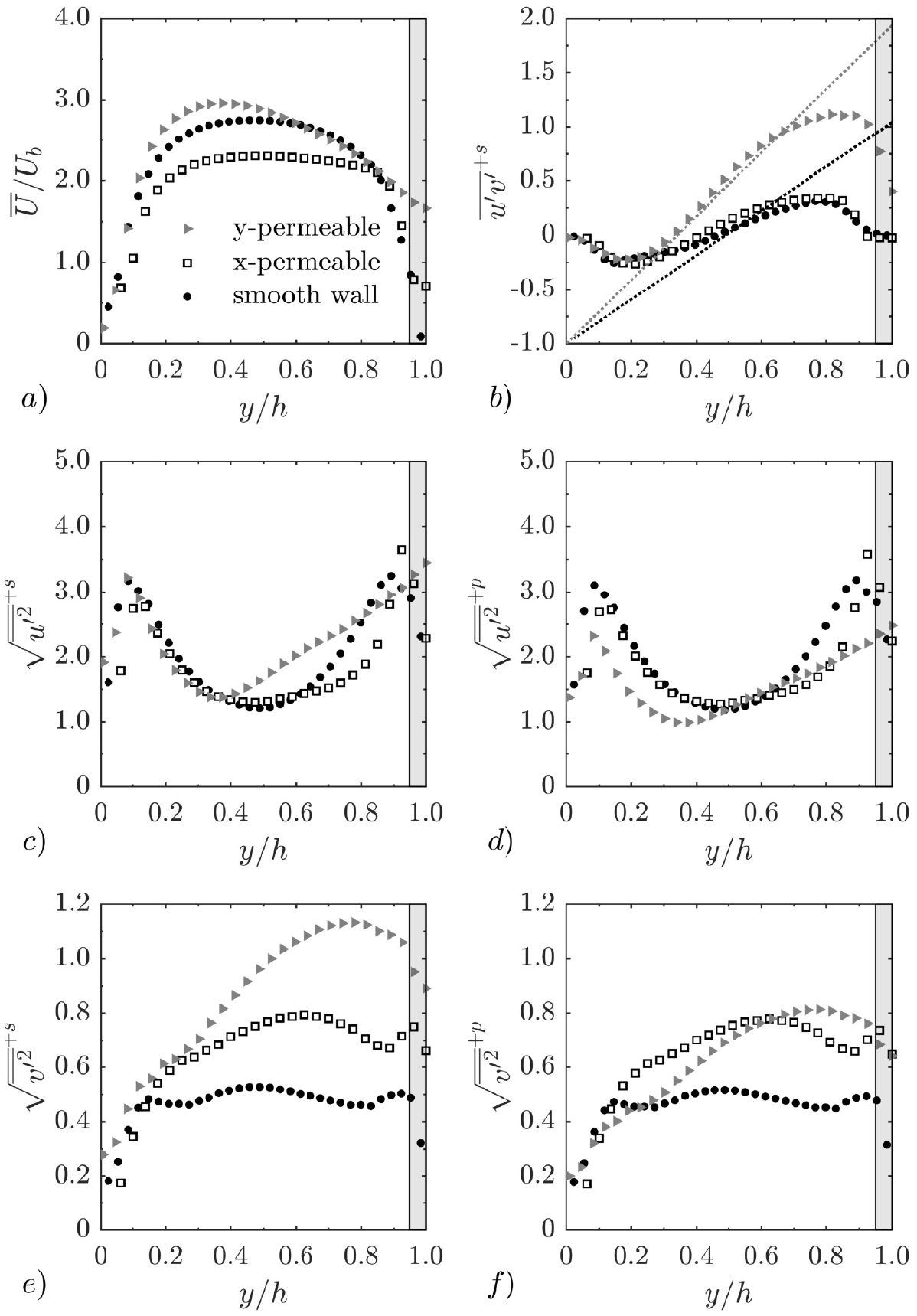}
	\caption{PIV results for the channel flow experiment. The smooth wall is located at $y=0$, while the interchangeable wall is located at $y=h$.  Panel (a) shows the measured mean velocity profiles normalized by the bulk-averaged velocity (calculated as $U_b = Q/[W(H+h)]$ in all cases). Panel (b) shows Reynolds shear stress profiles normalized by the smooth wall friction velocity $\protect\utaus$. 
	Panels (c,d) show profiles of root-mean-square streamwise fluctuations and panels (e,f) show profiles or root-mean-square wall-normal fluctuations. Panels (c,e) have been normalized by friction velocities at the smooth wall $\protect\utaus$, while panels (d,f) have been normalized by friction velocities at the porous interface, $\utaup$.} \label{fig:piv_results}
\end{figure*}

Figure~\ref{fig:piv_results} shows the measured mean statistics for the channel flow experiments. These statistics were computed by averaging both in time and in the streamwise direction. Results in the region $y/h\geq 0.95$ were affected by reflections at the smooth/porous tiles and should be treated with caution. 

As seen from Fig.~\ref{fig:piv_results}(a), the mean profile remains relatively symmetric across the unobstructed region for the x-permeable material. However, for the y-permeable case, the bulk of the flow in the unobstructed region is shifted towards the smooth wall. The location of the maximum mean velocity is $y/h = 0.45$ for the x-permeable case as compared to $y/h = 0.38$ for the y-permeable case.  Interestingly, the slip velocity at the porous interface appears to be higher for the y-permeable case despite the substantially lower streamwise permeability.  However, this observation could be attributed to the specific porous geometry tested here.  For the y-permeable material, the porous interface is characterized by much higher \textit{local} porosity compared to the x-permeable material.  The visibly lower bulk-normalized mean profile for the x-permeable material in the unobstructed region is indicative of greater flow through the porous medium itself. 

The Reynolds shear stress profiles in Fig.~\ref{fig:piv_results}(b) show the presence of an (almost) linear region in the middle of the unobstructed domain.  Friction velocities at the porous and smooth walls, $\utaup$ and $\utaus$, were estimated by extrapolating the total stress (i.e., Reynolds shear stress plus viscous stress) from this linear region to $y=h$ and $y=0$, respectively \citep[see e.g.,][]{breugem2006influence}. The viscous stress was estimated using velocity gradients computed from the measured mean velocity profiles. Near-wall velocity gradients estimated from the PIV measurements are expected to be inaccurate due to the relatively low wall-normal resolution ($\Delta y^+ \approx 3 - 5$).  As a result, friction velocities were estimated using a linear fit to the total stress only over the region between the maximum and minimum values of the Reynolds shear stress, i.e., measurements very close to the smooth wall and the porous interface were excluded. The resulting friction velocity estimates are listed in Table~\ref{table:exp_table}.  The average friction velocity was estimated as $u_\tau^t =\sqrt{((\utaus)^2+(\utaup)^2)/2}$.  Note that the profiles shown in Fig.~\ref{fig:piv_results}(b) are normalized by $\utaus$.  With this normalization, it is clear that $\utaup$ is higher than $\utaus$ for the y-permeable material, indicative of greater friction generated at the porous interface than at the smooth wall.  However, for the x-permeable material, the estimated friction velocities at the smooth wall and porous interface are comparable, $\utaus \approx \utaup$. In other words, there is no clear increase or decrease in friction at the porous interface relative to the smooth wall for the x-permeable material.  Friction velocities at both interfaces are approximately equal for the smooth wall case, as expected.  Thus, the friction velocity estimates are broadly consistent with the mean profiles shown in Fig.~\ref{fig:piv_results}(a); only the y-permeable material shows a significant difference in behavior at the porous interface. 

Moreover, relative to the smooth wall case, x-permeable material does not lead to a significant change in friction velocities.  Table~\ref{table:exp_table} shows that $\utaus$ values differ by 3\% and, at the porous wall, $\utaup$ is higher by approximately 4\% (note that $\utaup$ for the smooth wall corresponds to the value at the smooth tile at $y/h = 1$). In contrast, the y-permeable material leads to a significant increase in friction velocities relative to the smooth wall case; $\utaus$ increases by $20\%$ and $\utaup$ by $70\%$.  Thus, in contrast to the resolvent-based predictions for the NW cycle, no reduction in friction is observed at the porous interface for the x-permeable material.  However, the y-permeable material leads to a \textit{substantial} increase in friction at both the smooth wall and the porous interface. This is qualitatively consistent with model predictions, which show a substantial increase in NW cycle gain as well as the emergence of high-gain spanwise rollers over the y-permeable material.

 
\begin{table}
	\centering
	\newcommand{\tstrut}{\rule{0pt}{2.6ex}}
	\begin{tabular}{c c c c}
		\toprule
		Case  & $\utaus$ [m/s] & $\utaup$ [m/s] & $u_\tau^t$ [m/s]\\ \midrule 
		smooth wall &  0.0194 &  0.0198* &  0.0196 \\\tstrut
		x-permeable &  0.0201 &  0.0205  &  0.0203 \\ \tstrut
		y-permeable &  0.0235 &  0.0326  &  0.0284  \\ 
		\bottomrule
	\end{tabular}
    \caption{Friction velocity estimates at the smooth wall ($\protect\utaus$) and porous interface ($\protect\utaup$). The average friction velocity is $\protect\utaut$. For the smooth wall case $\protect\utaup$ corresponds to the solid tile placed in the cutout.}
	\label{table:exp_table}
\end{table}

Profiles for root-mean-square (RMS) streamwise velocity fluctuations are plotted in Figs.~\ref{fig:piv_results}(c) and \ref{fig:piv_results}(d), normalized by $\utaus$ and $\utaup$, respectively. Similarly, profiles for RMS wall-normal fluctuations are presented in Figs.~\ref{fig:piv_results}(e) and \ref{fig:piv_results}(f). When normalized by $\utaus$, the near-wall peaks in the streamwise fluctuations collapse together for all cases on the smooth wall side.  Further, the normalized peak value for the fluctuations is $\sqrt{\overbar{u^{\prime 2}}}^{+ s} \approx 3$, which is close to that expected in a canonical turbulent channel flow configuration. Figure~\ref{fig:piv_results}(d) shows that the peaks in streamwise RMS velocities for the smooth wall and x-permeable case collapse together reasonably well near the porous interface when normalized by $\utaup$, with a peak value of  $\sqrt{\overbar{u^{\prime 2}}}^{+ p} \approx 3$. This confirms that the flow physics are qualitatively similar for the smooth wall and x-permeable substrate.  However, the profile for the y-permeable case has no discernible peak near the porous interface and the maximum value for the normalized fluctuations  $\sqrt{\overbar{u^{\prime 2}}}^{+ p} \approx 2.5$, is lower than the other two cases. Figure~\ref{fig:piv_results}(e) shows that, when normalized by $\utaus$, the intensity of the wall-normal velocity fluctuations near the porous interface increases for both the x-permeable and y-permeable material relative to the smooth-wall case. However, this increase is much more pronounced for the y-permeable material. These observations are broadly consistent with previous results for flow over porous materials \citep{breugem2006influence,suga2010effects,suga2018anisotropic,kuwata2019extensive} and are indicative of a significant change in flow structure over the y-permeable material. 
	
\subsection{Flow Structure}\label{sec:POD}

\begin{figure*}[ht]
	\centering
	\includegraphics[width=1.0\textwidth]{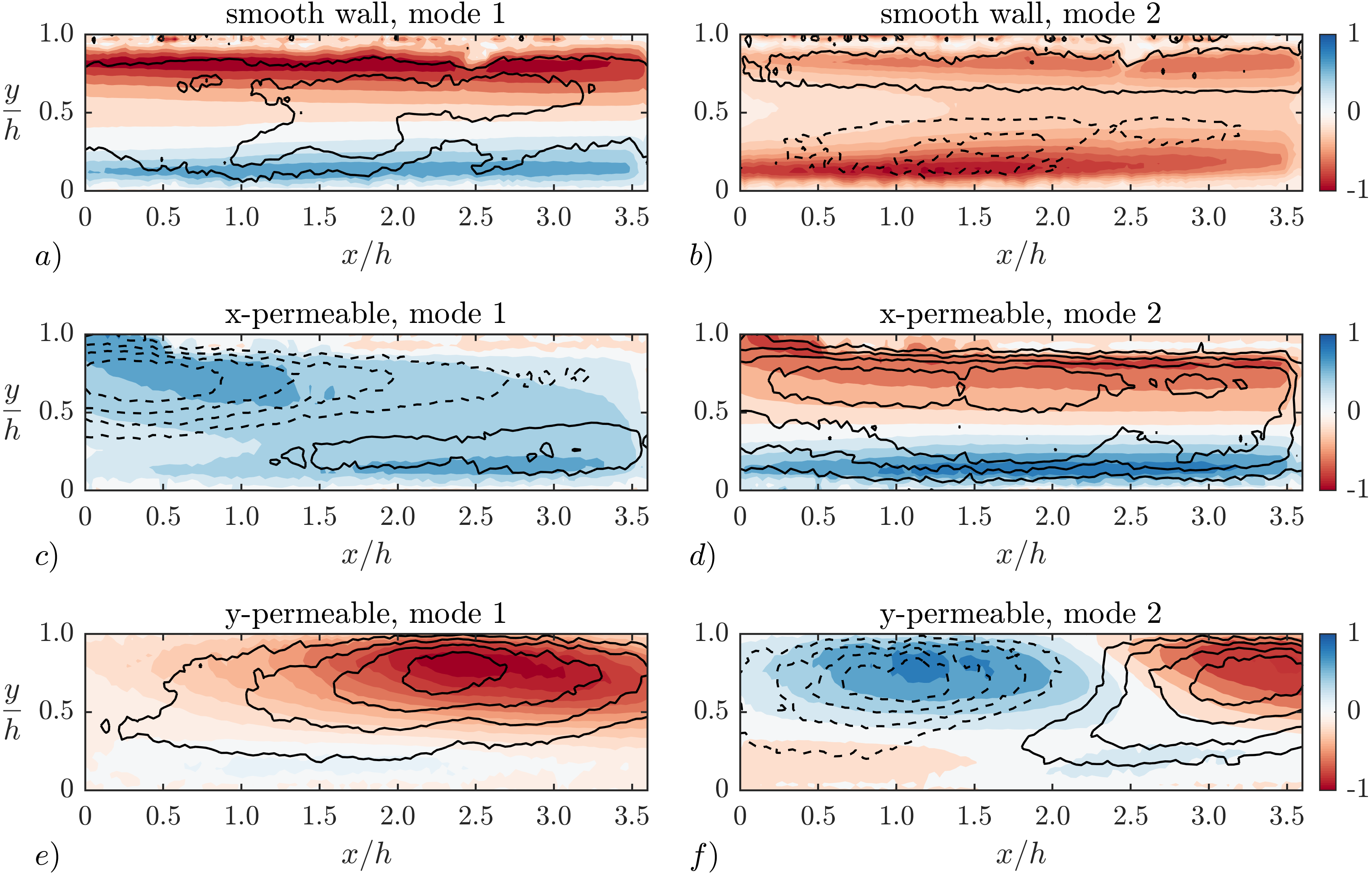}
	\caption{The first two POD modes for the smooth wall case (a,b), x-permeable case (c,d), and the y-permeable case (e,f). The modes are computed using 20,000 PIV frames. The shading represents normalized levels for the streamwise velocity component. The solid and dashed black lines represent positive and negative contours for the wall-normal velocity component. The porous interface is at the top wall.}
	\label{fig:pod}
\end{figure*}
	
To provide further insight into the changes in mean statistics discussed above, snapshot proper orthogonal decomposition (POD) was performed on the fluctuating velocity fields obtained from PIV. The streamwise velocity fields associated with the first 2 spatial modes for the smooth wall, x-permeable, and y-permeable cases are shown in Fig.~\ref{fig:pod}. 

As expected, the most energetic modes for the smooth wall case shown in Figs.~\ref{fig:pod}(a,b) resemble long streaky structures that are symmetric across the channel. The first POD mode for the x-permeable case shown in Fig.~\ref{fig:pod}(c) does not have a clear physical interpretation. It bears some resemblance to the second POD mode for the smooth wall shown in Fig.~\ref{fig:pod}(b) but is also similar to the asymmetric first POD mode over the y-permeable case shown in Fig.~\ref{fig:pod}(e). However, the second POD mode for the x-permeable substrate closely resembles the first mode for the smooth wall case.  The streamwise extent of the plots shown in Fig.~\ref{fig:pod} corresponds to the PIV field of view, which is roughly $3.5h$ or 22 mm.  With this in mind, the first smooth wall mode and the second x-permeable mode appear to have a streamwise wavelength that is more than twice the PIV field of view, $\lambda_x > 44$ mm (or $\lambda_x > 7h$). For the friction velocity estimates shown in Table~\ref{table:exp_table}, this translates into $\lambda_x^+ = (\lambda_x \utaut/ \nu) > 800$, which is consistent with the scale of NW streaks \citep{robinson1991coherent}.  

Unlike the smooth wall and x-permeable cases, POD modes for the y-permeable material have a visibly asymmetric structure in the wall-normal direction (see Fig.~\ref{fig:pod}(e,f)). For both modes, the streamwise velocity field is much more intense near the porous interface. This is in contrast to the symmetric streaky structures observed over both the smooth wall and x-permeable material. Moreover, when the streamwise and wall-normal velocity contours are considered together, the full velocity field for these POD modes is indicative of a counter-rotating structure. Such rollers have been observed in previous numerical simulations \citep{breugem2006influence,gomez-de-segura_garcia-mayoral_2019}, and are typically associated with a Kelvin-Helmholtz instability mechanism. This observation provides qualitative support for the model predictions shown in Fig.~\ref{fig:sigma_kh_combined}(d) which shows that the y-permeable material is susceptible to the emergence of large high-gain spanwise constant structures.  

More quantitatively, the first POD mode over the y-permeable material appears to have a streamwise wavelength more than twice the PIV field of view, $\lambda_x > 7h \approx 44$ mm.  Using the $\utaut$ estimate for the y-permeable material shown in Table~\ref{table:exp_table}, this translates into $\lambda_x^+ > 1200$.  Thus, the size of this structure is larger than the region of peak amplification around $\lambda_x^+ \approx 1000$ predicted in Fig.~\ref{fig:sigma_kh_combined}(d).  However, assuming that the spanwise rollers scale in outer units, $\lambda_x^+ \approx 1000$ for the predictions shown in Fig.~\ref{fig:sigma_kh_combined}(d) at $\Ret \approx 120$, corresponds to $\lambda_x/h = \lambda_x^+/\Ret \approx 8$. This is more consistent with the streamwise wavelength of the first POD mode for the y-permeable substrate shown in Fig.~\ref{fig:pod}(e). Put another way, if the resolvent-predictions had been carried out at the measured friction Reynolds number for the y-permeable substrate, $\Ret = \utaut h/\nu \approx 180$ for the $\utaut$ estimate shown in Table~\ref{table:exp_table}, they might show a region of higher amplification at higher streamwise wavelengths. This issue is explored further in \S\ref{sec:mean_profile}.

The streamwise extent of the second POD mode over the y-permeable substrate (see Fig.~\ref{fig:pod}(f)) is slightly larger than the PIV window, $\lambda_x \approx 4.5h$.  This corresponds to a streamwise wavelength of $\lambda_x^+ \approx 800$, which is at the lower end of the predicted region of highly-amplified spanwise rollers in Fig.~\ref{fig:sigma_kh_combined}(d).



\subsection{Model Sensitivity to Mean Profile and Permeability}\label{sec:mean_profile}

\begin{figure*}[ht]
	\centering
	\includegraphics[width =1.0\textwidth]{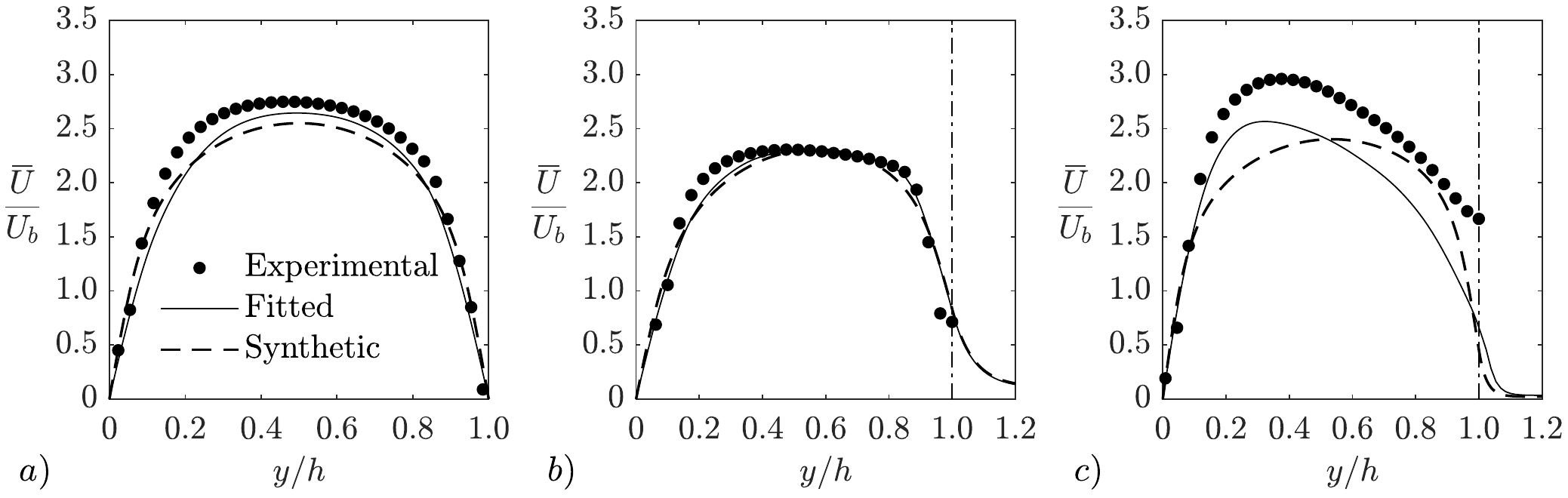}
	\caption{Comparison of the mean velocity profiles used in the construction of the resolvent operator for the smooth wall (a), x-permeable material (b), and y-permeable material (c). Experimental profiles are shown using the circular makers (\protect\bcircle).  Synthetic profiles generated using the eddy viscosity model of \citet{reynolds_tiederman_1967} are plotted as dashed lines (\protect\bdash). Fitted profiles computed using eddy viscosity profile determined from experimental data are plotted as solid lines (\protect\bline).}
	\label{fig:mean_vel_comparison}
\end{figure*}

The results presented in the previous sections show that there is no clear evidence of NW-cycle suppression and friction reduction over the x-permeable material.  For the y-permeable material, a substantial increase in friction is observed at the porous interface relative to the smooth wall and x-permeable material.  In addition, the POD modes shown in \S\ref{sec:POD} suggest that large-scale spanwise rollers are energetic over the y-permeable substrate.  These results are in partial agreement with the \textit{a priori} model predictions shown in \S\ref{sec:model_predictions}.  Here, we revisit resolvent-based predictions for the NW mode and spanwise rollers, but using mean profiles estimated from the experiments based on the procedure described in \S\ref{sec:implementation} and measured permeability values. 

Figure~\ref{fig:mean_vel_comparison} compares mean velocity profiles measured in the experiments (\bcircle) with those generated using the synthetic eddy viscosity profile \citep{reynolds_tiederman_1967} (\bdash) and the fitted eddy viscosity profile (\bline).  In general, the synthetic and fitted profiles are in close agreement with the measurements for the smooth wall and x-permeable material, as shown in Figs.~\ref{fig:mean_vel_comparison}(a,b). However, Fig.~\ref{fig:mean_vel_comparison}(c) shows that the synthetic mean profile for the y-permeable case does not reproduce the asymmetry observed in the experiments.  This is because the synthetic eddy viscosity profile from \citet{reynolds_tiederman_1967} was developed for smooth wall flows.  Therefore, it assumes that the turbulence is symmetric across the unobstructed channel. This is reasonable for the smooth wall case and the x-permeable material but not for the y-permeable materials (see $\utaus$ and $\utaup$ estimates in Table~\ref{table:exp_table}). The fitted eddy viscosity profile accounts for the asymmetry in turbulence across the unobstructed region over the y-permeable substrate.  As a result, it is able to better reproduce the shape of the measured mean profile. Note that the eddy viscosity models lead to a marginal under-prediction in bulk-normalized mean velocity profiles for the smooth wall and x-permeable cases. This could be attributed to spanwise variation in the mean flow in the finite-width channel (recall that the channel has an aspect ratio of $W/h \approx 8$).  The mean profile measured at the channel centerline may be slightly larger than the true spanwise average.  However, the eddy viscosity model leads to a more significant under-prediction in the mean velocity profile for the y-permeable material. We do not have a clear explanation for this under-prediction.

\begin{table}
	\centering
	\newcommand{\tstrut}{\rule{0pt}{2.6ex}}
	\begin{tabular}{ccc}
    \toprule
	& {predicted $\bfK$} & {measured $\bfK$} \\ \cmidrule(r){2-3}
	&  synthetic profile  & fitted profile \\ 
	\midrule\tstrut
	x-permeable, $\sigma_{\bfk,p}/\sigma_{\bfk,s}$ &  0.71  &  0.75  \\ 
	y-permeable, $\sigma_{\bfk,p}/\sigma_{\bfk,s}$ &  2.55 & 1.87 \\
	\bottomrule
	\end{tabular}
	\caption{Comparison of the normalized gain for NW resolvent modes predicted using \textit{a priori} information (numerically predicted $\bfK$ and synthetic mean profile) and quantities estimated from laboratory measurements (measured $\bfK$ and fitted mean profile).}
	\label{table:nw_vals}
\end{table}

Table~\ref{table:nw_vals} shows how the normalized gain for the NW resolvent mode changes with the mean profile and permeability for the x-permeable and y-permeable substrates.  These normalized singular values are computed for friction Reynolds numbers estimated from the experiments: $\Ret = \utaut h/\nu \approx 124$ for the smooth wall case, $\Ret \approx 129$ for the x-permeable substrate, and $\Ret \approx 180$ for the y-permeable substrate.  In other words, the singular values over the porous substrates ($\sigma_{\bfk,p}$) are normalized by the singular values for a smooth wall ($\sigma_{\bfk,s}$) at the $\Ret$ estimated from the experiments.  For these values of $\Ret$, the measured permeability values listed in Table~\ref{table:perm_table} translate into $(\sqrt{\kxps}, \sqrt{\kyps}) \approx (8.2,1.4)$ for the x-permeable material and $(\sqrt{\kxps}, \sqrt{\kyps}) \approx (1.9,11.4)$ for the y-permeable material.  For reference, the \textit{a priori} values for these quantities used in \S\ref{sec:model_predictions} were $(\sqrt{\kxps}, \sqrt{\kyps}) \approx (7.8,2.9)$ for the x-permeable material and $(\sqrt{\kxps}, \sqrt{\kyps}) \approx (2.9,7.8)$ for the y-permeable material.

For the x-permeable material, the singular value ratio increases from $\sigma_{\bfk,p}/\sigma_{\bfk,s}$ = $0.71$ for the synthetic mean profile and predicted permeability to $\sigma_{\bfk,p}/\sigma_{\bfk,s} = 0.75$ for the fitted mean profile and measured permeability.  In other words, the predicted suppression for the NW resolvent mode decreases from $29\%$ to $25\%$.  Despite this slight deterioration, the model still predicts suppression for the NW mode, which we consider a necessary, but not sufficient, condition for friction reduction.  For the y-permeable substrate, the singular value ratio decreases from  $\sigma_{\bfk,p}/\sigma_{\bfk,s} = 2.55$ for the synthetic profile and predicted permeability to $\sigma_{\bfk,p}/\sigma_{\bfk,s} = 1.87$ for the fitted profile and the measured permeability.  Thus, model predictions made using the measured quantities yield more limited NW mode amplification ($87\%$) compared to the \textit{a priori} predictions ($155\%$).  The measured increase in the average friction velocity for the y-permeable material relative to the smooth wall case is approximately $45\%$ (see $\utaut$ values in Table~\ref{table:exp_table}), which is close to half the predicted increase in NW mode amplification.  However, it must be stressed that the change in NW mode amplification is not a direct measure of drag reduction performance.  The x-permeable material yields a $4\%$ increase in $\utaut$ relative to the smooth wall case, even though the resolvent model predicts suppression for the NW mode. 

Together, the predictions shown in Table~\ref{table:nw_vals} indicate that: (i) the exact permeability value and the shape of the mean profile can have a significant effect on resolvent-based predictions, and (ii) the eddy viscosity model from \citet{reynolds_tiederman_1967} may not be the most appropriate choice for generating mean profile predictions over porous substrates.  However, the overarching design guidelines do not change: only materials with high streamwise permeability and low wall-normal/spanwise permeabilities are likely to \ml{suppress the NW resolvent mode and, potentially,} reduce drag. 

\begin{figure*}[ht]
	\centering
	\includegraphics[width=0.8\textwidth]{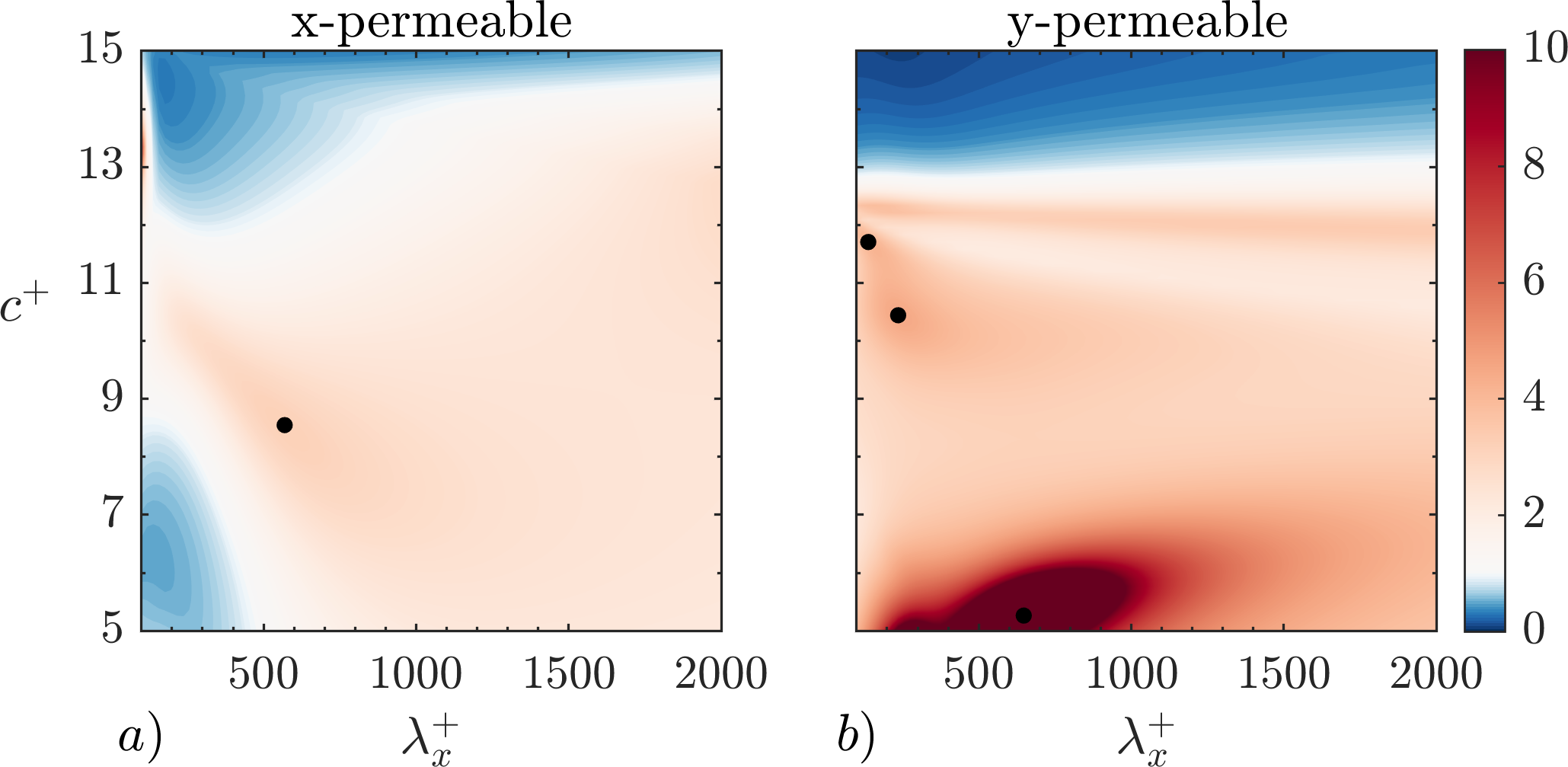}
	\caption{Normalized gain ($\sigma_{\bfk,p}/\sigma_{\bfk,s}$) for spanwise-constant resolvent modes as a function of streamwise wavelength and mode speed computed using the measured permeability values shown in Table~\ref{table:perm_table} and mean velocity profiles fitted to the experimental measurements.  Panel (a) shows predictions for the x-permeable material while panel (b) shows predictions for the y-permeable material.}
	\label{fig:sigma_kh_compare}
\end{figure*}

\begin{figure*}[ht]
	\centering
	\includegraphics[width=1.0\textwidth]{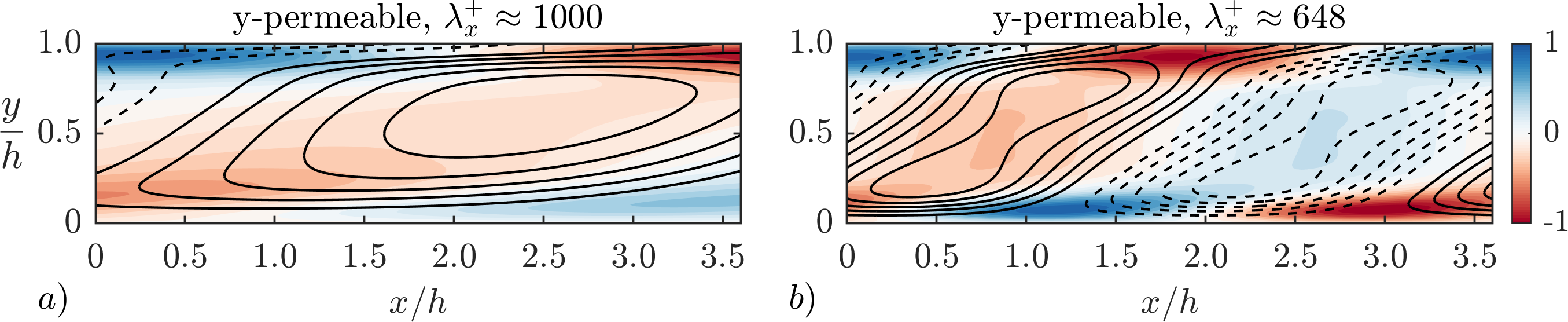}
	\caption{High-gain spanwise-constant resolvent modes identified from model predictions for the y-permeable substrate.  (a) Structure corresponding to the highest-gain mode from the synthetic mean profile predictions in Fig.~\ref{fig:sigma_kh_combined}(d) with $(\lambda_x^+,c^+) \approx (1000,8.5)$. (b) Structure corresponding to the highest-gain mode from the fitted mean profile predictions in Fig.~\ref{fig:sigma_kh_compare}(d) with $(\lambda_x^+,c^+) \approx (650,5)$. The shading represents normalized levels for the streamwise velocity component. Solid and dashed black lines respectively represent positive and negative contours for the wall-normal velocity component. The porous interface is at the top wall ($y/h=1$).}
	\label{fig:resolvent_HG_modes}
\end{figure*}

Next, we evaluate how resolvent-based predictions for the spanwise-constant modes change with the mean profile and permeability. For the x-permeable case, there is no localized region of very high amplification for either the \textit{a priori} predictions shown in Fig.~\ref{fig:sigma_kh_combined}(c) or the predictions based on measured quantities shown in Fig.~\ref{fig:sigma_kh_compare}(a).  However, for the y-permeable case, the new predictions in Fig.~\ref{fig:sigma_kh_compare}(b) show some important changes relative to the model predictions presented earlier in Fig.~\ref{fig:sigma_kh_combined}(d). Specifically, the localized peak in amplification observed in Fig.~\ref{fig:sigma_kh_combined}(d) around $(\lambda_x^+, c^+) \approx (1000, 8.5)$ shifts to a region extending from $(\lambda_x^+,c^+) \approx (300,5)$ to $(\lambda_x^+,c^+) \approx (1000,6)$ in Fig.~\ref{fig:sigma_kh_compare}(b).  It could be argued that the second POD mode for the y-permeable substrate shown in Fig.~\ref{fig:pod}(f), which is estimated to have a streamwise extent of $\lambda_x^+ \approx 800$, belongs in this region.  However, there is no clear evidence of a longer structure like resembling the first POD mode shown in Fig.~\ref{fig:pod}(e).


Finally, we present resolvent-based predictions for flow structure over the y-permeable material. The predicted structure for spanwise-constant modes with the highest normalized gain in Fig.~\ref{fig:sigma_kh_combined}(d) and Fig.~\ref{fig:sigma_kh_compare}(b) is shown in Fig.~\ref{fig:resolvent_HG_modes}(a) and Fig.~\ref{fig:resolvent_HG_modes}(b), respectively.  The streamwise extent of these structures is roughly consistent with the second POD mode for the y-permeable material shown in Fig.~\ref{fig:pod}(f).  Moreover, both predicted mode shapes show a structure that is inclined across the channel rather than being symmetric or anti-symmetric. This observation also agrees qualitatively with the shape of the second POD mode in Fig.~\ref{fig:pod}(f).  However, there are significant differences in flow structure between the predictions and the measurements.  Specifically, the streamwise velocity field in the predicted mode shapes is localized primarily near the wall and the porous interface.  In contrast, the POD mode shows a much larger wall-normal extent for the streamwise velocity field.  This discrepancy is the source of ongoing research. 

\section{Conclusion}\label{sec:conclusion}
Consistent with previous theoretical efforts and numerical simulations, the present experiments show a very different flow response over the x-permeable and y-permeable materials. The x-permeable material leads to a marginal increase in friction velocity at the porous interface (see Fig.~\ref{fig:piv_results}(b) and Table~\ref{table:exp_table}). This is counter to resolvent-based predictions, which suggest that the x-permeable material should lead to a reduction in gain for the NW cycle (Fig.~\ref{fig:sigma_kh_combined}(a)). Possible explanations for this discrepancy include: the emergence of energetic spanwise rollers, as predicted by the resolvent formulation and previous numerical simulations for materials with $\sqrt{\kyps}>0.4$; roughness effects at the porous interface that are neglected in the VANS equations;  nonlinear (Forchheimer) effects becoming important in the porous medium; and perhaps insufficient scale separation between the pore-scale and outer flow. The y-permeable material leads to a significant increase in friction velocities at both walls relative to the smooth wall and x-permeable cases. This is consistent with resolvent-based predictions, which indicate a substantial increase in NW cycle gain as well as the emergence of large, high-gain spanwise rollers over the y-permeable material. POD confirms the presence of such spanwise rollers over the y-permeable material (Fig.~\ref{fig:pod}).  Resolvent-based predictions are able to predict the streamwise length scale of the POD mode. However, there are important differences in structure between the highest-gain resolvent modes and the computed POD modes.

The model predictions and experimental results shown here confirm that materials with high streamwise permeability and low spanwise and wall-normal permeability are good candidates for drag reduction. However, it must be emphasized again that the preliminary experiments discussed in this study do not show drag reduction. Moreover, they only consider a single low Reynolds number and do not explicitly evaluate the effect of the interfacial roughness. Indeed, decoupling the effects of permeability and interfacial roughness is an important issue that needs to be addressed in future studies.  
Ongoing work seeks to alleviate some of the weaknesses associated with the current experimental setup (limited development length, insufficient scale separation, low Reynolds number) and to identify streamwise-permeable porous materials that could be more effective.

\section{Acknowledgements}
This material is based on work supported by the Air Force Office of Scientific Research under awards FA9550-17-1-0142 (program manager Dr. Gregg Abate) and FA9550-19-1-7027 (program manager Dr. Douglas Smith).  We also thank three anonymous reviewers for providing valuable feedback on this manuscript.

\bibliography{ac_ce_ml}

\begin{thebibliography}{56}
\expandafter\ifx\csname natexlab\endcsname\relax\def\natexlab#1{#1}\fi
\providecommand{\url}[1]{\texttt{#1}}
\providecommand{\href}[2]{#2}
\providecommand{\path}[1]{#1}
\providecommand{\DOIprefix}{doi:}
\providecommand{\ArXivprefix}{arXiv:}
\providecommand{\URLprefix}{URL: }
\providecommand{\Pubmedprefix}{pmid:}
\providecommand{\doi}[1]{\href{http://dx.doi.org/#1}{\path{#1}}}
\providecommand{\Pubmed}[1]{\href{pmid:#1}{\path{#1}}}
\providecommand{\bibinfo}[2]{#2}
\ifx\xfnm\relax \def\xfnm[#1]{\unskip,\space#1}\fi
\bibitem[{Luchini et~al.(1991)Luchini, Manzo, and
  Pozzi}]{luchini1991resistance}
\bibinfo{author}{P.~Luchini}, \bibinfo{author}{F.~Manzo},
  \bibinfo{author}{A.~Pozzi}, \bibinfo{journal}{Journal of Fluid Mechanics}
  \bibinfo{volume}{228} (\bibinfo{year}{1991}) \bibinfo{pages}{87--109}.
\bibitem[{Robert(1992)}]{robert1992drag}
\bibinfo{author}{J.~Robert}, \bibinfo{title}{Drag reduction: an industrial
  challenge}, \bibinfo{type}{Technical Report}, AIRBUS INDUSTRIE BLAGNAC
  (FRANCE), \bibinfo{year}{1992}.
\bibitem[{Walsh and Lindemann(1984)}]{walsh1984optimization}
\bibinfo{author}{M.~Walsh}, \bibinfo{author}{A.~Lindemann}, in:
  \bibinfo{booktitle}{22nd Aerospace Sciences Meeting}, p.
  \bibinfo{pages}{347}.
\bibitem[{Garcia-Mayoral and Jimenez(2011)}]{garcia2011hydrodynamic}
\bibinfo{author}{R.~Garcia-Mayoral}, \bibinfo{author}{J.~Jimenez},
  \bibinfo{journal}{Journal of Fluid Mechanics} \bibinfo{volume}{678}
  (\bibinfo{year}{2011}) \bibinfo{pages}{317--347}.
\bibitem[{Robinson(1991)}]{robinson1991coherent}
\bibinfo{author}{S.~K. Robinson}, \bibinfo{journal}{Annual Review of Fluid
  Mechanics} \bibinfo{volume}{23} (\bibinfo{year}{1991})
  \bibinfo{pages}{601--639}.
\bibitem[{Smits et~al.(2011)Smits, McKeon, and Marusic}]{smits2011high}
\bibinfo{author}{A.~J. Smits}, \bibinfo{author}{B.~J. McKeon},
  \bibinfo{author}{I.~Marusic}, \bibinfo{journal}{Annual Review of Fluid
  Mechanics} \bibinfo{volume}{43} (\bibinfo{year}{2011})
  \bibinfo{pages}{353--375}.
\bibitem[{Choi et~al.(1993)Choi, Moin, and Kim}]{choi1993direct}
\bibinfo{author}{H.~Choi}, \bibinfo{author}{P.~Moin}, \bibinfo{author}{J.~Kim},
  \bibinfo{journal}{Journal of Fluid Mechanics} \bibinfo{volume}{255}
  (\bibinfo{year}{1993}) \bibinfo{pages}{503--539}.
  \DOIprefix\doi{10.1017/S0022112093002575}.
\bibitem[{Lee and Lee(2001)}]{lee2001flow}
\bibinfo{author}{S.-J. Lee}, \bibinfo{author}{S.-H. Lee},
  \bibinfo{journal}{Experiments in Fluids} \bibinfo{volume}{30}
  (\bibinfo{year}{2001}) \bibinfo{pages}{153--166}.
  \DOIprefix\doi{10.1007/s003480000150}.
\bibitem[{Abderrahaman-Elena and
  Garc\'{\i}a-Mayoral(2017)}]{nabil_garcia_dragreduction}
\bibinfo{author}{N.~Abderrahaman-Elena},
  \bibinfo{author}{R.~Garc\'{\i}a-Mayoral}, \bibinfo{journal}{Phys. Rev.
  Fluids} \bibinfo{volume}{2} (\bibinfo{year}{2017}) \bibinfo{pages}{114609}.
  \DOIprefix\doi{10.1103/PhysRevFluids.2.114609}.
\bibitem[{Rosti et~al.(2018)Rosti, Brandt, and Pinelli}]{rosti2018turbulent}
\bibinfo{author}{M.~E. Rosti}, \bibinfo{author}{L.~Brandt},
  \bibinfo{author}{A.~Pinelli}, \bibinfo{journal}{Journal of Fluid Mechanics}
  \bibinfo{volume}{842} (\bibinfo{year}{2018}) \bibinfo{pages}{381--394}.
\bibitem[{{G\'omez-de-Segura} and
  García-Mayoral(2019)}]{gomez-de-segura_garcia-mayoral_2019}
\bibinfo{author}{G.~{G\'omez-de-Segura}}, \bibinfo{author}{R.~García-Mayoral},
  \bibinfo{journal}{Journal of Fluid Mechanics} \bibinfo{volume}{875}
  (\bibinfo{year}{2019}) \bibinfo{pages}{124--172}.
  \DOIprefix\doi{10.1017/jfm.2019.482}.
\bibitem[{McKeon and Sharma(2010)}]{mckeon_sharma_2010_sharma_2010}
\bibinfo{author}{B.~J. McKeon}, \bibinfo{author}{A.~S. Sharma},
  \bibinfo{journal}{Journal of Fluid Mechanics} \bibinfo{volume}{658}
  (\bibinfo{year}{2010}) \bibinfo{pages}{336--382}.
  \DOIprefix\doi{10.1017/S002211201000176X}.
\bibitem[{Luhar et~al.(2014)Luhar, Sharma, and McKeon}]{luhar2014opposition}
\bibinfo{author}{M.~Luhar}, \bibinfo{author}{A.~S. Sharma},
  \bibinfo{author}{B.~J. McKeon}, \bibinfo{journal}{Journal of Fluid Mechanics}
  \bibinfo{volume}{749} (\bibinfo{year}{2014}) \bibinfo{pages}{597--626}.
\bibitem[{Luhar et~al.(2015)Luhar, Sharma, and McKeon}]{luhar2015framework}
\bibinfo{author}{M.~Luhar}, \bibinfo{author}{A.~S. Sharma},
  \bibinfo{author}{B.~McKeon}, \bibinfo{journal}{Journal of Fluid Mechanics}
  \bibinfo{volume}{768} (\bibinfo{year}{2015}) \bibinfo{pages}{415--441}.
\bibitem[{Chavarin and Luhar(2020)}]{chavarin2020resolvent}
\bibinfo{author}{A.~Chavarin}, \bibinfo{author}{M.~Luhar},
  \bibinfo{journal}{AIAA Journal} \bibinfo{volume}{58} (\bibinfo{year}{2020})
  \bibinfo{pages}{589--599}.
\bibitem[{García-Mayoral et~al.(2019)García-Mayoral, Gómez-de Segura, and
  Fairhall}]{gm_gs_fairhall_2019}
\bibinfo{author}{R.~García-Mayoral}, \bibinfo{author}{G.~Gómez-de Segura},
  \bibinfo{author}{C.~T. Fairhall}, \bibinfo{journal}{Fluid Dynamics Research}
  (\bibinfo{year}{2019}) \bibinfo{pages}{011410}.
\bibitem[{Busse and Sandham(2012)}]{busse_Sandham_2012}
\bibinfo{author}{A.~Busse}, \bibinfo{author}{N.~D. Sandham},
  \bibinfo{journal}{Physics of Fluids} \bibinfo{volume}{24}
  (\bibinfo{year}{2012}) \bibinfo{pages}{055111}.
  \DOIprefix\doi{10.1063/1.4719780}.
\bibitem[{G{\'o}mez-de Segura et~al.(2018)G{\'o}mez-de Segura, Sharma, and
  Garc{\'\i}a-Mayoral}]{gomez2018turbulent}
\bibinfo{author}{G.~G{\'o}mez-de Segura}, \bibinfo{author}{A.~Sharma},
  \bibinfo{author}{R.~Garc{\'\i}a-Mayoral}, \bibinfo{journal}{Flow, Turbulence
  and Combustion} \bibinfo{volume}{100} (\bibinfo{year}{2018})
  \bibinfo{pages}{995--1014}.
\bibitem[{Jim{\'e}nez and Pinelli(1999)}]{jimenez_pinelli_1999}
\bibinfo{author}{J.~Jim{\'e}nez}, \bibinfo{author}{A.~Pinelli},
  \bibinfo{journal}{Journal of Fluid Mechanics} \bibinfo{volume}{389}
  (\bibinfo{year}{1999}) \bibinfo{pages}{335--359}.
  \DOIprefix\doi{10.1017/S0022112099005066}.
\bibitem[{Breugem et~al.(2006)Breugem, Boersma, and
  Uittenbogaard}]{breugem2006influence}
\bibinfo{author}{W.~Breugem}, \bibinfo{author}{B.~Boersma},
  \bibinfo{author}{R.~Uittenbogaard}, \bibinfo{journal}{Journal of Fluid
  Mechanics} \bibinfo{volume}{562} (\bibinfo{year}{2006})
  \bibinfo{pages}{35--72}.
\bibitem[{Rosti et~al.(2015)Rosti, Cortelezzi, and Quadrio}]{rosti2015direct}
\bibinfo{author}{M.~E. Rosti}, \bibinfo{author}{L.~Cortelezzi},
  \bibinfo{author}{M.~Quadrio}, \bibinfo{journal}{Journal of Fluid Mechanics}
  \bibinfo{volume}{784} (\bibinfo{year}{2015}) \bibinfo{pages}{396--442}.
\bibitem[{Kuwata and Suga(2016)}]{kuwata2016lattice}
\bibinfo{author}{Y.~Kuwata}, \bibinfo{author}{K.~Suga},
  \bibinfo{journal}{International Journal of Heat and Fluid Flow}
  \bibinfo{volume}{61} (\bibinfo{year}{2016}) \bibinfo{pages}{145 -- 157}.
  \DOIprefix\doi{https://doi.org/10.1016/j.ijheatfluidflow.2016.03.006}.
\bibitem[{Kuwata and Suga(2017)}]{kuwata2017direct}
\bibinfo{author}{Y.~Kuwata}, \bibinfo{author}{K.~Suga},
  \bibinfo{journal}{Journal of Fluid Mechanics} \bibinfo{volume}{831}
  (\bibinfo{year}{2017}) \bibinfo{pages}{41--71}.
\bibitem[{Kuwata and Suga(2019)}]{kuwata2019extensive}
\bibinfo{author}{Y.~Kuwata}, \bibinfo{author}{K.~Suga},
  \bibinfo{journal}{International Journal of Heat and Fluid Flow}
  \bibinfo{volume}{80} (\bibinfo{year}{2019}) \bibinfo{pages}{108465}.
\bibitem[{Zagni and Smith(1976)}]{zagni1976channel}
\bibinfo{author}{A.~F. Zagni}, \bibinfo{author}{K.~V. Smith},
  \bibinfo{journal}{Journal of the Hydraulics Division} \bibinfo{volume}{102}
  (\bibinfo{year}{1976}) \bibinfo{pages}{207--222}.
\bibitem[{Pokrajac and Manes(2009)}]{pokrajac2009velocity}
\bibinfo{author}{D.~Pokrajac}, \bibinfo{author}{C.~Manes},
  \bibinfo{journal}{Transport in Porous Media} \bibinfo{volume}{78}
  (\bibinfo{year}{2009}) \bibinfo{pages}{367}.
  \DOIprefix\doi{10.1007/s11242-009-9339-8}.
\bibitem[{Horton and Pokrajac(2009)}]{horton2009onset}
\bibinfo{author}{N.~Horton}, \bibinfo{author}{D.~Pokrajac},
  \bibinfo{journal}{Physics of Fluids} \bibinfo{volume}{21}
  (\bibinfo{year}{2009}) \bibinfo{pages}{045104}.
\bibitem[{Kim et~al.(2016)Kim, Blois, Best, and
  Christensen}]{kim2016experimental}
\bibinfo{author}{T.~Kim}, \bibinfo{author}{G.~Blois},
  \bibinfo{author}{J.~Best}, \bibinfo{author}{K.~T. Christensen}, in:
  \bibinfo{booktitle}{River Flow 2016}, \bibinfo{publisher}{CRC Press},
  \bibinfo{year}{2016}, pp. \bibinfo{pages}{950--955}.
\bibitem[{Suga et~al.(2010)Suga, Matsumura, Ashitaka, Tominaga, and
  Kaneda}]{suga2010effects}
\bibinfo{author}{K.~Suga}, \bibinfo{author}{Y.~Matsumura},
  \bibinfo{author}{Y.~Ashitaka}, \bibinfo{author}{S.~Tominaga},
  \bibinfo{author}{M.~Kaneda}, \bibinfo{journal}{International Journal of Heat
  and Fluid Flow} \bibinfo{volume}{31} (\bibinfo{year}{2010})
  \bibinfo{pages}{974--984}.
\bibitem[{Manes et~al.(2011)Manes, Poggi, and Ridolfi}]{manes2011turbulent}
\bibinfo{author}{C.~Manes}, \bibinfo{author}{D.~Poggi},
  \bibinfo{author}{L.~Ridolfi}, \bibinfo{journal}{Journal of Fluid Mechanics}
  \bibinfo{volume}{687} (\bibinfo{year}{2011}) \bibinfo{pages}{141--170}.
\bibitem[{Efstathiou and Luhar(2018)}]{efstathiou2018mean}
\bibinfo{author}{C.~Efstathiou}, \bibinfo{author}{M.~Luhar},
  \bibinfo{journal}{Journal of Fluid Mechanics} \bibinfo{volume}{841}
  (\bibinfo{year}{2018}) \bibinfo{pages}{351--379}.
\bibitem[{Kim et~al.(2020)Kim, Blois, Best, and
  Christensen}]{kim2020experimental}
\bibinfo{author}{T.~Kim}, \bibinfo{author}{G.~Blois}, \bibinfo{author}{J.~L.
  Best}, \bibinfo{author}{K.~T. Christensen}, \bibinfo{journal}{Journal of
  Fluid Mechanics} \bibinfo{volume}{887} (\bibinfo{year}{2020}).
\bibitem[{Kong and Schetz(1982)}]{kong1982turbulent}
\bibinfo{author}{F.~Kong}, \bibinfo{author}{J.~Schetz}, in:
  \bibinfo{booktitle}{20th Aerospace Sciences Meeting}, p.~\bibinfo{pages}{30}.
\bibitem[{Suga et~al.(2018)Suga, Okazaki, Ho, and Kuwata}]{suga2018anisotropic}
\bibinfo{author}{K.~Suga}, \bibinfo{author}{Y.~Okazaki},
  \bibinfo{author}{U.~Ho}, \bibinfo{author}{Y.~Kuwata},
  \bibinfo{journal}{Journal of Fluid Mechanics} \bibinfo{volume}{855}
  (\bibinfo{year}{2018}) \bibinfo{pages}{983--1016}.
\bibitem[{Suga et~al.(2020)Suga, Okazaki, and Kuwata}]{suga2020characteristics}
\bibinfo{author}{K.~Suga}, \bibinfo{author}{Y.~Okazaki},
  \bibinfo{author}{Y.~Kuwata}, \bibinfo{journal}{Journal of Fluid Mechanics}
  \bibinfo{volume}{884} (\bibinfo{year}{2020}).
\bibitem[{Itoh et~al.(2006)Itoh, Tamano, Iguchi, Yokota, Akino, Hino, and
  Kubo}]{itoh2006turbulent}
\bibinfo{author}{M.~Itoh}, \bibinfo{author}{S.~Tamano},
  \bibinfo{author}{R.~Iguchi}, \bibinfo{author}{K.~Yokota},
  \bibinfo{author}{N.~Akino}, \bibinfo{author}{R.~Hino},
  \bibinfo{author}{S.~Kubo}, \bibinfo{journal}{Physics of Fluids}
  \bibinfo{volume}{18} (\bibinfo{year}{2006}) \bibinfo{pages}{065102}.
\bibitem[{Moarref et~al.(2013)Moarref, Sharma, Tropp, and
  McKeon}]{moarref2013model}
\bibinfo{author}{R.~Moarref}, \bibinfo{author}{A.~S. Sharma},
  \bibinfo{author}{J.~A. Tropp}, \bibinfo{author}{B.~J. McKeon},
  \bibinfo{journal}{Journal of Fluid Mechanics} \bibinfo{volume}{734}
  (\bibinfo{year}{2013}) \bibinfo{pages}{275--316}.
  \DOIprefix\doi{10.1017/jfm.2013.457}.
\bibitem[{Nakashima et~al.(2017)Nakashima, Fukagata, and
  Luhar}]{nakashima2017assessment}
\bibinfo{author}{S.~Nakashima}, \bibinfo{author}{K.~Fukagata},
  \bibinfo{author}{M.~Luhar}, \bibinfo{journal}{Journal of Fluid Mechanics}
  \bibinfo{volume}{828} (\bibinfo{year}{2017}) \bibinfo{pages}{496--526}.
\bibitem[{Toedtli et~al.(2019)Toedtli, Luhar, and
  McKeon}]{toedtli2019predicting}
\bibinfo{author}{S.~S. Toedtli}, \bibinfo{author}{M.~Luhar},
  \bibinfo{author}{B.~J. McKeon}, \bibinfo{journal}{Physical Review Fluids}
  \bibinfo{volume}{4} (\bibinfo{year}{2019}) \bibinfo{pages}{073905}.
\bibitem[{Zampogna and Bottaro(2016)}]{zampogna_bottaro_2016}
\bibinfo{author}{G.~A. Zampogna}, \bibinfo{author}{A.~Bottaro},
  \bibinfo{journal}{Journal of Fluid Mechanics} \bibinfo{volume}{792}
  (\bibinfo{year}{2016}) \bibinfo{pages}{5--35}.
  \DOIprefix\doi{10.1017/jfm.2016.66}.
\bibitem[{Luhar et~al.(2014)Luhar, Sharma, and McKeon}]{luhar2014structure}
\bibinfo{author}{M.~Luhar}, \bibinfo{author}{A.~Sharma},
  \bibinfo{author}{B.~McKeon}, \bibinfo{journal}{Journal of Fluid Mechanics}
  \bibinfo{volume}{751} (\bibinfo{year}{2014}) \bibinfo{pages}{38--70}.
\bibitem[{McKeon(2017)}]{mckeon2017engine}
\bibinfo{author}{B.~McKeon}, \bibinfo{journal}{Journal of Fluid Mechanics}
  \bibinfo{volume}{817} (\bibinfo{year}{2017}).
\bibitem[{Aurentz and Trefethen(2017)}]{rectangular_block_operators_trefethen}
\bibinfo{author}{J.~Aurentz}, \bibinfo{author}{L.~Trefethen},
  \bibinfo{journal}{SIAM Review} \bibinfo{volume}{59} (\bibinfo{year}{2017})
  \bibinfo{pages}{423--446}. \DOIprefix\doi{10.1137/16M1065975}.
\bibitem[{Ochoa-Tapia and Whitaker(1995)}]{ochoa1995momentum1}
\bibinfo{author}{J.~A. Ochoa-Tapia}, \bibinfo{author}{S.~Whitaker},
  \bibinfo{journal}{International Journal of Heat and Mass Transfer}
  \bibinfo{volume}{38} (\bibinfo{year}{1995}) \bibinfo{pages}{2635--2646}.
  \DOIprefix\doi{10.1016/0017-9310(94)00346-W}.
\bibitem[{Tilton and Cortelezzi(2006)}]{tilton_cortelezzi_2006}
\bibinfo{author}{N.~Tilton}, \bibinfo{author}{L.~Cortelezzi},
  \bibinfo{journal}{Physics of Fluids} \bibinfo{volume}{18}
  (\bibinfo{year}{2006}) \bibinfo{pages}{051702}.
  \DOIprefix\doi{10.1063/1.2202649}.
\bibitem[{Tilton and Cortelezzi(2008)}]{tilton_cortelezzi_2008}
\bibinfo{author}{N.~Tilton}, \bibinfo{author}{L.~Cortelezzi},
  \bibinfo{journal}{Journal of Fluid Mechanics} \bibinfo{volume}{604}
  (\bibinfo{year}{2008}) \bibinfo{pages}{411--445}.
  \DOIprefix\doi{10.1017/S0022112008001341}.
\bibitem[{Reynolds and Tiederman(1967)}]{reynolds_tiederman_1967}
\bibinfo{author}{W.~C. Reynolds}, \bibinfo{author}{W.~G. Tiederman},
  \bibinfo{journal}{Journal of Fluid Mechanics} \bibinfo{volume}{27}
  (\bibinfo{year}{1967}) \bibinfo{pages}{253--–272}.
  \DOIprefix\doi{10.1017/S0022112067000308}.
\bibitem[{Vinuesa et~al.(2014)Vinuesa, Noorani, Lozano-Dur{\'a}n, Khoury,
  Schlatter, Fischer, and Nagib}]{vinuesa2014aspect}
\bibinfo{author}{R.~Vinuesa}, \bibinfo{author}{A.~Noorani},
  \bibinfo{author}{A.~Lozano-Dur{\'a}n}, \bibinfo{author}{G.~K.~E. Khoury},
  \bibinfo{author}{P.~Schlatter}, \bibinfo{author}{P.~F. Fischer},
  \bibinfo{author}{H.~M. Nagib}, \bibinfo{journal}{Journal of Turbulence}
  \bibinfo{volume}{15} (\bibinfo{year}{2014}) \bibinfo{pages}{677--706}.
\bibitem[{Thielicke and Stamhuis(2014)}]{thielicke2014pivlab}
\bibinfo{author}{W.~Thielicke}, \bibinfo{author}{E.~J. Stamhuis},
  \bibinfo{journal}{Journal of Open Research Software} \bibinfo{volume}{2}
  (\bibinfo{year}{2014}). \URLprefix
  \url{http://openresearchsoftware.metajnl.com/articles/10.5334/jors.bl/}.
  \DOIprefix\doi{10.5334/jors.bl}.
\bibitem[{Byrne et~al.(1969)Byrne, Hatton, and Marriott}]{byrne1969turbulent}
\bibinfo{author}{J.~Byrne}, \bibinfo{author}{A.~Hatton},
  \bibinfo{author}{P.~Marriott}, \bibinfo{journal}{Proceedings of the
  Institution of Mechanical Engineers} \bibinfo{volume}{184}
  (\bibinfo{year}{1969}) \bibinfo{pages}{697--712}.
\bibitem[{Zanoun et~al.(2009)Zanoun, Kito, and Egbers}]{zanoun2009study}
\bibinfo{author}{E.-S. Zanoun}, \bibinfo{author}{M.~Kito},
  \bibinfo{author}{C.~Egbers}, \bibinfo{journal}{Journal of Fluids Engineering}
  \bibinfo{volume}{131} (\bibinfo{year}{2009}).
  \DOIprefix\doi{10.1115/1.3112384}.
\bibitem[{Lien et~al.(2004)Lien, Monty, Chong, and Ooi}]{lien2004entrance}
\bibinfo{author}{K.~Lien}, \bibinfo{author}{J.~P. Monty},
  \bibinfo{author}{M.~S. Chong}, \bibinfo{author}{A.~Ooi}, in:
  \bibinfo{booktitle}{15th Australian Fluid Mechanics Conference},
  volume~\bibinfo{volume}{15}, pp. \bibinfo{pages}{356--363}.
\bibitem[{Vinuesa et~al.(2014)Vinuesa, Bartrons, Chiu, Dressler, R{\"u}edi,
  Suzuki, and Nagib}]{vinuesa2014new}
\bibinfo{author}{R.~Vinuesa}, \bibinfo{author}{E.~Bartrons},
  \bibinfo{author}{D.~Chiu}, \bibinfo{author}{K.~M. Dressler},
  \bibinfo{author}{J.-D. R{\"u}edi}, \bibinfo{author}{Y.~Suzuki},
  \bibinfo{author}{H.~M. Nagib}, \bibinfo{journal}{Experiments in Fluids}
  \bibinfo{volume}{55} (\bibinfo{year}{2014}) \bibinfo{pages}{1759}.
\bibitem[{Antonia and Luxton(1971)}]{antonia1971response}
\bibinfo{author}{R.~Antonia}, \bibinfo{author}{R.~Luxton},
  \bibinfo{journal}{Journal of Fluid Mechanics} \bibinfo{volume}{48}
  (\bibinfo{year}{1971}) \bibinfo{pages}{721--761}.
\bibitem[{Le et~al.(1997)Le, Moin, and Kim}]{le1997direct}
\bibinfo{author}{H.~Le}, \bibinfo{author}{P.~Moin}, \bibinfo{author}{J.~Kim},
  \bibinfo{journal}{Journal of Fluid Mechanics} \bibinfo{volume}{330}
  (\bibinfo{year}{1997}) \bibinfo{pages}{349--374}.
\bibitem[{Chandesris et~al.(2013)Chandesris, D'Hueppe, Mathieu, Jamet, and
  Goyeau}]{chandesris2013direct}
\bibinfo{author}{M.~Chandesris}, \bibinfo{author}{A.~D'Hueppe},
  \bibinfo{author}{B.~Mathieu}, \bibinfo{author}{D.~Jamet},
  \bibinfo{author}{B.~Goyeau}, \bibinfo{journal}{Physics of Fluids}
  \bibinfo{volume}{25} (\bibinfo{year}{2013}) \bibinfo{pages}{125110}.

\end{thebibliography}
	
\end{document}